\documentclass[apj]{emulateapj}
\usepackage{epstopdf}
\usepackage{url}

\def\crot{\citet{Crockett:11:78}}
\def\maht{\citet{Mahmud:11:123}}
\def\prat{\citet{Prato:08:L103}}
\def\crop{\citep{Crockett:11:78}}
\def\mahp{\citep{Mahmud:11:123}}
\def\prap{\citep{Prato:08:L103}}

\def\mjup{\hbox{$M_{\textrm{\scriptsize{JUP}}}$}}
\def\kmps{\hbox{km s${}^{-1}$}}
\def\mps{\hbox{m s${}^{-1}$}}
\def\msd{\hbox{m s${}^{-1}$ d${}^{-1}$}}
\def\teff{\hbox{$T_{\textrm{\scriptsize{eff}}}$}}
\def\deg{\hbox{$^\circ$}}

\makeatletter
\newcommand\footnoteref[1]{\protected@xdef\@thefnmark{\ref{#1}}\@footnotemark}
\makeatother

\shorttitle{A Search for Giant Planet Companions to T Tauri Stars}
\shortauthors{Crockett et al.}

\slugcomment{Accepted to ApJ: November 5, 2012}
\begin{document}

\title{A search for giant planet companions to T Tauri stars}
\author{Christopher J. Crockett \altaffilmark{1,3}, \\
	Naved I. Mahmud \altaffilmark{3,4}, \\
	L. Prato \altaffilmark{2,3}, \\
	Christopher M. Johns-Krull \altaffilmark{3,4}, \\	
	Daniel T. Jaffe \altaffilmark{5}, \\
        Patrick M. Hartigan \altaffilmark{4}, \\
	Charles A. Beichman \altaffilmark{6,7}}

\altaffiltext{1}{U.S. Naval Observatory, 10391 W. Naval Observatory Road, Flagstaff, AZ 86001; ccrockett@nofs.navy.mil}
\altaffiltext{2}{Lowell Observatory, 1400 W Mars Hill Road, Flagstaff, AZ 86001; lprato@lowell.edu}
\altaffiltext{3}{Visiting Astronomer at the Infrared Telescope Facility, which is operated by the University of Hawaii under Cooperative Agreement no. NCC 5-538 with the National Aeronautics and Space Administration, Science Mission Directorate, Planetary Astronomy Program.}
\altaffiltext{4}{Department of Physics and Astronomy, Rice University, MS-108, 6100 Main Street, Houston, TX 77005; naved@rice.edu, cmj@rice.edu}
\altaffiltext{5}{Department of Astronomy, University of Texas, R.L. Moore Hall, Austin, TX 78712; dtj@astro.as.utexas.edu}
\altaffiltext{6}{Jet Propulsion Laboratory, California Institute of Technology, 4800 Oak Grove Drive, Pasadena, CA 91109}
\altaffiltext{7}{NASA Exoplanet Science Institute (NExScI), California Institute of Technology, 770 S. Wilson Ave, Pasadena, CA 91125}

\begin{abstract}
We present results from an ongoing multiwavelength radial velocity (RV) survey of the Taurus-Auriga star forming region as part of our effort to identify pre--main sequence giant planet hosts.  These 1--3 Myr old T Tauri stars present significant challenges to traditional RV surveys.  The presence of strong magnetic fields gives rise to large, cool star spots.  These spots introduce significant RV jitter which can mimic the velocity modulation from a planet-mass companion.  To distinguish between spot-induced and planet-induced RV modulation, we conduct observations at $\sim$6700 \AA\ and $\sim$2.3 \micron\ and measure the wavelength dependence (if any) in the RV amplitude.  CSHELL observations of the known exoplanet host Gl 86 demonstrate our ability to detect not only hot Jupiters in the near infrared but also secular trends from more distant companions.  Observations of nine very young stars reveal a typical reduction in RV amplitude at the longer wavelengths by a factor of $\sim$2--3.  While we can not confirm the presence of planets in this sample, three targets show different periodicities in the two wavelength regions.  This suggests different physical mechanisms underlying the optical and K band variability. 
\end{abstract}



\section{Introduction}
The discovery of over 840 exoplanets\footnote{\label{note1}\url{http://exoplanet.eu}} in the past twenty years has revealed that planetary systems are common and diverse. Pulsar planets \citep{Wolszczan:94:538}, hot Jupiters \citep{Mayor:95:355}, and retrograde orbits \citep{Hebrard:11:L11} repeatedly challenge our assumptions about planet formation. After two decades of exoplanet discoveries, the processes underlying planet formation remain unclear. Lacking direct observational inputs, theorists must deduce formation mechanisms from observations of mature systems.  A more preferable approach is to catalog the planet population around very young ($\sim$1 Myr old) stars while these nascent worlds still reside in their formation environments.  

Unfortunately, observations of pre--main sequence stars are complicated by star spots, jets, accretion, and circumstellar disks. The most successful attempts at young planet detection to date have come from direct imaging surveys.  For example, \citet{Lafreniere:10:497} presented evidence for a giant planet around the  $\sim$5 Myr old Upper Sco solar analogue 1RXS J160929.1--210524.  They report the presence of an $\sim$8 \mjup\ companion at a distance of 330 AU based on JHK adaptive optics images from NIRI on Gemini North.  Another direct imaging detection is reported in \citet{Kraus:12:5}.  They present evidence for a faint point source $\sim$16 AU from the 2 Myr old LkCa 15 using non-redundant aperture masking interferometry on Keck II.  They intepret this source as a $\sim$6 \mjup\ protoplanet. These observations, and others like them, are providing the first glimpse into the mechanisms of giant planet formation.

There are, however, limitations to what can currently be accomplished with direct imaging.  This technique can only detect massive planets at wide separation from their host stars.  Furthermore, imaging does not provide a direct measure of companion mass; the derived masses are model-dependent.  Radial velocity (RV) surveys, which have confirmed $\sim$95\% of the known exoplanet population\footnoteref{note1}, obtain mass limits and are more suited to identifying planets on close orbits. In particular, ``hot Jupiters'' are readily accessible via RV observations.

Low-mass young stars, however, are challenging targets for traditional RV surveys. These targets are faint at optical wavelengths owing to late spectral types, large distances ($>$100 pc), and extinction from natal dust clouds.  Young stars also have strong magnetic fields \citep[e.g.,][]{Johns-Krull:07:975} that generate large, cool star spots. Spots impact RV surveys of young stars by introducing significant jitter which can mimic the RV modulation induced by a planet \citep{Saar:97:319, Hatzes:02:392, Desort:07:983, Reiners:10:432}.  Recent attempts at detecting substellar companions in young stellar populations (10--100 Myr) via RV monitoring have generally been unsuccessful \citep[e.g.,][]{Paulson:04:3579, Paulson:06:706}, likely because of small sample sizes and the intrinsic RV variability of the targets.  For very young, 1--3 Myr old stars, the RV noise problem is even more acute.  Recently, \citet{vanEyken:12} reported observations of a possible hot Jupiter orbiting a 2--3 Myr old weak-lined T Tauri star in Orion ($P \approx 0.45$ days).  This interpretation is based on transit data from the Palomar Transit Factory (PTF) project in conjunction with optical RV observations from the Hobby-Eberly Telescope (HET) and Keck I. While this is an intriguing result, their data are dominated by spot modulation.  Confirmation must wait for further observations, particularly in the near infrared (NIR).  The youngest confirmed RV planet to date is a $\sim$6 \mjup\ planet on a 850 day orbit around the 100 Myr old star HD 70573 \citep{Setiawan:07:L145}.

The success of a young star RV survey depends on its ability to distinguish between spot-induced and companion-induced RV modulation.  Spot-induced RV variations in young stars have been identified by correlations between RVs and spectral line bisector spans \citep[e.g.,][]{Hatzes:97:374}; planet hosts do not exhibit any such correlation.  However, there may also be no correlation if the absorption lines are not spectroscopically resolved, i.e., the projected rotation velocity ($v$ sin $i$) of the star is comparable to or less than the velocity resolution of the spectrograph \citep{Desort:07:983, Huelamo:08:L9, Prato:08:L103}. A potentially more reliable method for distinguishing between spots and planets leverages the wavelength dependence of a spot-induced RV modulation amplitude.

The reflex motion caused by a planet affects all wavelengths equally.  However, the amplitude of any spot-induced RV variability will be smaller at longer wavelengths \citep[e.g.,][]{Huelamo:08:L9, Prato:08:L103, Reiners:10:432, Ma:12:172}.  This wavelength dependence arises because of the flux-temperature scaling in the Rayleigh-Jeans limit of blackbody radiation; the contrast between a photosphere and a cooler star spot decreases at longer wavelengths \citep[e.g.,][]{Vrba:86:199, Carpenter:01:3160}. Therefore, observations in the visible \emph{and} NIR may distinguish between stellar activity and a true companion by comparing the RV amplitudes at the two wavelengths. Furthermore, the late spectral types of T Tauri stars produce SEDs with peak emission around 1--2 \micron~and extinction of these partly embedded sources is lessened at longer wavelengths. NIR observations can therefore lead to increased signal-to-noise and improved RV precision.

NIR spectroscopy has only recently begun to play a role in the search for substellar companions.  \citet{Martin:06:L75} combined optical and H band observations to look for giant planets around the young brown dwarf LP 944-20. They concluded that the observed optical RV modulations were driven by inhomogeneous surface features (i.e. clouds).  \citet{Blake:07:1198} used high-resolution K band spectroscopy to investigate the presence of giant planets around a population of L dwarfs, achieving a precision of 300 \mps.  They found no evidence for companions with $M \sin i > 2$ \mjup\ and $P < 3$ days in their sample of nine targets.  \citet{Setiawan:08:38} reported the optical detection of a $\sim$10 \mjup\ planet on a 3.5 day orbit around the 10 Myr old star TW Hydra. However, H band observations by \citet{Huelamo:08:L9} revealed a strong wavelength dependence in the velocity amplitude.  The low H band variability casts doubt on the presence of a companion and suggests that spots cause TW Hydra's optical RV variations.  \citet{Prato:08:L103} observed DN Tau and V836 Tau---two potential planet-host candidates selected on the basis of their optical variability---in the K band.  To within their measurement precision, they detected no NIR RV modulation, implying that spots produce the apparent optical RV variability.  

Interest in targeting M dwarfs as hosts of habitable planets has spurred vast improvements in NIR precision, bringing it closer to that of optical surveys.  \citet{Bean:10:410} report a long term precision in the H band of $\sim$5 \mps\ on late M dwarfs using an ammonia gas cell on CRIRES at the VLT.  \citet{Figueira:10:55} report a comparable precision, also on CRIRES, but using telluric absorption features as a wavelength reference. Based on six years of NIRSPEC data, \citet{Blake:10:684} report a precision of 50 \mps\ for their sample of K dwarfs and 200 \mps\ for L dwarfs also through the use of telluric lines.  \citet{Bailey:12:16} also report 50 \mps\ precision on mid-M dwarfs, increasing to $\sim$80--170 \mps\ for young ($\sim$10 Myr old) stars, using NIRSPEC with telluric-based wavelength calibration. While these results are encouraging, these efforts are all focused on large (8--10 meter) aperture telescopes.  Very little work has been done to develop precision NIR RV techniques for smaller aperture telescopes where more observing time is available to the community.  We recently presented our technique for achieving $\sim$50 \mps\ on late-K and early-M dwarfs using the CSHELL NIR spectrograph on the IRTF \crop.  \citet{Anglada-Escude:12} also show very promising results with their successful efforts at obtaining 20--30 \mps\ precision in the K band, also on CSHELL, using a methane isotopologue gas cell.

Since 2004, we have been conducting an RV survey in the Taurus-Auriga star forming region in an attempt to catalog the giant planet and brown dwarf population around very young stars \citep{Huerta:08:472, Prato:08:L103, Crockett:11:78, Mahmud:11:123}. In this paper, we present new and updated analyses on nine targets including known spotted stars and potential young planet hosts; the entire sample will be discussed in a later paper once the entire survey is complete.  In \S\ref{sec:observing} we provide an overview of our observing strategy.  Our data reduction pipeline and RV analysis techniques are presented in \S\ref{sec:reduction}, including results from RV standards used to quantify our long-term precision.  In \S\ref{sec:rvanalysis}, we discuss the results of these analyses for each of our T Tauri targets, as well as updated results on the known exoplanet host Gl 86.  Finally, in \S\ref{sec:discussion} we discuss the relevance of these results and how they can inform future studies of the young planet population.

\begin{deluxetable*}{lcccccccc}
\tablecaption{Stellar parameters of young planet host candidates and spotted star standards}
\tablewidth{350pt}
\tablehead{
\multicolumn{1}{c}{Name} & \multicolumn{1}{c}{V\tablenotemark{a}} &  \multicolumn{1}{c}{K\tablenotemark{a}} & \multicolumn{1}{c}{\teff\tablenotemark{b}} & \multicolumn{1}{c}{$v \sin i$\tablenotemark{c}} & \multicolumn{1}{c}{Type\tablenotemark{b}} & \multicolumn{1}{c}{Age\tablenotemark{b}} \\
\multicolumn{1}{c}{} & \multicolumn{1}{c}{(mag)} &  \multicolumn{1}{c}{(mag)} & \multicolumn{1}{c}{(K)} & \multicolumn{1}{c}{(\kmps)} & \multicolumn{1}{c}{(C/W)\tablenotemark{d}} & \multicolumn{1}{c}{(Myr)} }
\startdata
BP Tau   & 12.3 & 7.7 & 4060 &  10.9 & C & 1.8 \\
CI Tau   & 13.0 & 7.8 & 4060 &  10.6 & C & 2.1 \\
DK Tau   & 12.6 & 7.1 & 4060 &  11.5 & C & 1.0 \\
DN Tau   & 12.5 & 8.0 & 3850 &   8.7 & C & 0.9 \\
Hubble I 4 & 12.0 & 7.3 & 4060 &  12.9 & W & 0.7 \\
IQ Tau   & 14.5 & 7.8 & 3780 &  12.0 & C & 1.4 \\
IW Tau   & 12.4 & 8.3 & 4060 &   6.9 & W & 2.1 \\
V827 Tau & 12.2 & 8.2 & 4060 &  19.3 & W & 2.5 \\
V836 Tau & 13.1 & 8.6 & 4060 &  15.0 & C & 7.9 \\
\enddata
\tablenotetext{a}{SIMBAD}
\tablenotetext{b}{\citet{Palla:02:1194}}
\tablenotetext{c}{\citet{Glebocki:05}}
\tablenotetext{d}{C = classical, W = weak-lined}
\label{tbl:stellarparameters}
\end{deluxetable*}

\section{Observations}
\label{sec:observing}
Our target list mainly consists of very young (1--3 Myr) T Tauri stars in the nearby \citep[$\sim$140 pc,][]{Kenyon:08:405} Taurus-Auriga star forming region, many of which we have been observing for several years.  These targets are selected from \citet{Herbig:88} based on the following criteria: (1) V $<$ 15, (2) $v \sin i <$ 20 \kmps\ to ensure relatively narrow, deep photospheric lines for accurate RV analysis, and (3) GKM stars to maximize S/N in the red part of the spectrum for our infrared observations.  Pre--main sequence stars of late K and early M spectral types will evolve to become late G, early K solar analogues.  They offer a good compromise between having deep photospheric CO lines needed for NIR RV measurements (\S\ref{sec:NIRRVs}) and being massive enough to offer a reasonable chance of hosting a giant planet \citep{Johnson:10:905}.  Of the 129 T Tauri stars, $\sim$35\% are classical (actively accreting) and $\sim$65\% are weak-lined (little to no accretion).  We also include 14 targets from the older ($\sim$100 Myr) Pleiades open cluster.  About 20\% of our sample are known to be close ($\gtrsim$ 0\arcsec.05) binaries which are unresolved in our observations.  Although these targets are RV variables, the timescale is on the order of tens to hundreds of years and thus does not impact our search for short-period variability.

We conduct observations in both the optical ($\sim$6200--7100 \AA) and the K band ($\sim$2.3 \micron) to identify any wavelength dependence on RV variability.  Our observing strategy starts with obtaining optical observations of a subset of targets every night for roughly a week at a time.  A week-long observing window matches the typical rotation period of T Tauri stars and falls within the range of companion orbital periods we are capable of detecting.  On the basis of the optical data, we then group our observed targets into two categories: (1) stars that exhibit correlations between RV variations and line bisector changes, and (2) stars that show no such correlations. The RV variations of stars in Group 1 are likely attributable to star spots \citep{Huerta:08:472} while stars in Group 2 may harbor substellar companions.  However, since a lack of RV-bisector correlation is not sufficient for definite identification of a companion, we test the planet hypothesis by conducting follow-up observations of Group 2 stars in the NIR \citep{Prato:08:L103,Mahmud:11:123}.  

With this strategy, we only target stars that show no correlation between RV and bisector span for further observations in the NIR.  The question then arises as to how many RV-bisector pairs are needed to identify a correlation between these two quantities for a typical spotted T Tauri star. To address this question, we examined randomly sampled subsets of various sizes from the RV and bisector data of the spotted young star LkCa 19 \citep{Huerta:08:472}. Using Monte Carlo methods, we found a correlation between the RV and bisector data 90\% of the time at a false alarm probability (FAP) of 0.05 when we sampled twelve observations. Hence, with twelve measurements per star, 1 in 20 legitimate exoplanet systems will be erroneously rejected by this test and about 1 in 10 spotted stars with clear correlations will mistakenly pass through to the next stage of observations.  While more observations per star would improve these numbers, twelve observations strikes a reasonable balance between the need to have a large sample of stars and good time-coverage per star.

\subsection{McDonald Observatory}
\label{obs_mcdonald}
Visible light spectra were taken at the McDonald Observatory 2.7 meter Harlan J. Smith telescope on the Robert G. Tull Coud\'{e} Spectrograph \citep{Tull:95:251}. The cross-dispersed coud\'{e} covers a wavelength range of 3986--9952 \AA.  We isolated the orders with strongest S/N, 6233--7109 \AA, for RV analysis.   A 1\arcsec.2 slit yielded a spectral resolving power of $R\equiv\frac{\lambda}{\Delta\lambda}\approx60,000$. Integration times were typically 1800 s (depending on conditions) and typical seeing was $\sim$2\arcsec. We took Thorium-Argon (ThAr) lamp exposures before and after each target observation for wavelength calibration; typical RMS values for the dispersion solution were $\sim$4 \mps\ \mahp.  

\subsection{IRTF}
\label{sec:obs_cshell}

The majority of our NIR observations were obtained at the 3 meter NASA Infrared Telescope Facility (IRTF) using CSHELL \citep{Tokunaga:90:131, Greene:93:313}.   CSHELL is a long-slit echelle spectrograph (1.08--5.5 $\mu$m) that uses a Circular Variable Filter (CVF) to isolate a single order onto a 256$\times$256 InSb detector array.  We used the CVF to isolate a 50 \AA~segment of spectrum centered at 2.298 $\mu$m (order 25).  This region contains numerous deep photospheric absorption lines from the 2.293 \micron\  CO $\nu$ = 2--0 band head as well a rich set of predominately CH$_4$ telluric absorption features which we use as a wavelength reference.  The 0\arcsec.5 slit yielded a typical FWHM of 2.6 pixels (0.5 \AA, 6.5 \kmps, measured from arc lamp spectra) corresponding to a spectral resolving power of R $\sim$ 46,000.   

We obtained data on 56 nights between Feb 2008 and Feb 2012.  These observations included several RV standards, a known exoplanet host, and nine T Tauri planet host candidates (Table \ref{tbl:stellarparameters}).  At the beginning of each night, we imaged twenty flat fields, each with a 20-second integration time, using a continuum lamp to illuminate the entire slit.  We also imaged the same number of 20-second dark frames.  Additionally, we imaged six Ar-Kr-Xe emission lines, changing the CVF while maintaining the grating position, to determine the wavelength reference.  All of our target data were obtained using 10\arcsec~nodded pairs to enable subtraction of sky emission, dark current, and detector bias.  Integration times for each nod were typically 600 seconds; for fainter targets we took multiple contiguous nod pairs \crop

\subsection{Keck}
\label{sec:obs_nirspec}
We supplemented our CSHELL data with six nights of observations using NIRSPEC \citep{McLean:98:566, McLean:00:1048} between Mar 2007 and Jan 2012. NIRSPEC is a NIR, vacuum-cryogenic, high-resolution, cross-dispersed, echelle spectrograph operating at the Nasymth focus on the 10 meter Keck II telescope.  We used the N7 filter (1.839--2.630 \micron) with the echelle and cross-disperser angles set to 62.72\deg\ and 36.24\deg, respectively, to image orders 30 through 35 onto the 1024 $\times$ 1024 InSb detector.  This provided a wavelength coverage of roughly 2.157--2.550 \micron.  The 0.288\arcsec\ $\times$ 24\arcsec\ slit yielded a median FWHM of 2.25 pixels (0.74 \AA, 9.6 \kmps, measured from arc lamp spectra) corresponding to a spectral resolving power of R $\approx$ 31,000.  To optimize simultaneous telluric and photosphere line coverage (\S\ref{sec:NIRRVs}), we limited subsequent analyses to order 33 (2.286--2.320 \micron).

\begin{deluxetable}{lr}
\tablecaption{Optical RV - bisector correlations}
\tablewidth{100pt}
\tablehead{
\colhead{Target} & \colhead{$r$}}
\startdata
BP Tau &  0.20 \\
CI Tau &  0.74 \\
DK Tau &  0.60 \\
DN Tau &  0.33 \\
Hubble 4 &  0.80 \\
IQ Tau &  0.28 \\
IW Tau & -0.16 \\
V827 Tau & -0.88 \\
V836 Tau & -0.30 \\
\enddata
\label{tbl:rvbis}
\end{deluxetable}

We observed RV standards, telluric standards, and over a dozen potential planet hosts in Taurus.  At the beginning of each night, we imaged approximately ten flat fields (3 coadds $\times$ 30 sec) using a continuum lamp to illuminate the entire slit.  We also imaged the same number of dark frames (3 coadds $\times$ 30 sec).  Additionally, we imaged a Ne-Ar-Xe-Kr calibration lamp to provide an initial rough estimate of the wavelength reference.  All of our target data were obtained using an ABBA 10\arcsec\ nod sequence to enable subtraction of sky emission, dark current, and detector bias.  Integration times were typically 30--60 seconds with 2--3 coadds.

\section{Data reduction}
\label{sec:reduction}
\subsection{Visible light radial velocities}
\label{sec:VLRVs}

All spectra were bias-subtracted, flat-fielded, and optimally extracted using an IDL echelle reduction code \citep{Hinkle:00}.  We determined visible light RVs with a cross-correlation analysis of eight echelle orders, each covering about 100 \AA. The orders were chosen for their high S/N, lack of stellar emission lines, and lack of telluric absorption lines.  For each target's cross-correlation template spectrum, we chose the highest S/N observation.  Using the target itself as a template reduces additional uncertainty from any spectral type mismatching.  Since each order was cross-correlated against the corresponding order in the template, our RVs are measured relative to one observation epoch.  For each observation, the RV and its associated uncertainty were taken to be the mean and standard deviations, respectively, of the RVs from all utilized echelle orders.  

To determine our long-term RV uncertainty, we observed several stars known to be stable at a few \mps: 107 Psc, HD 4628, $\tau$ Ceti, HD 65277, HD 80367, and HD 88371 \citep{Nidever:02:503, Butler:96:500, Cumming:99:890}.  The standard deviation of these RV standards over eight years is $\sim$140 \mps.  We adopt this value as our long-term systematic uncertainty.  The final uncertainty for each observation is then 140 \mps\ added in quadrature with the order-to-order scatter \mahp.

\subsection{Visible light bisector analysis}
\label{sec:VLBis}

\begin{figure*}
\includegraphics[scale=0.65]{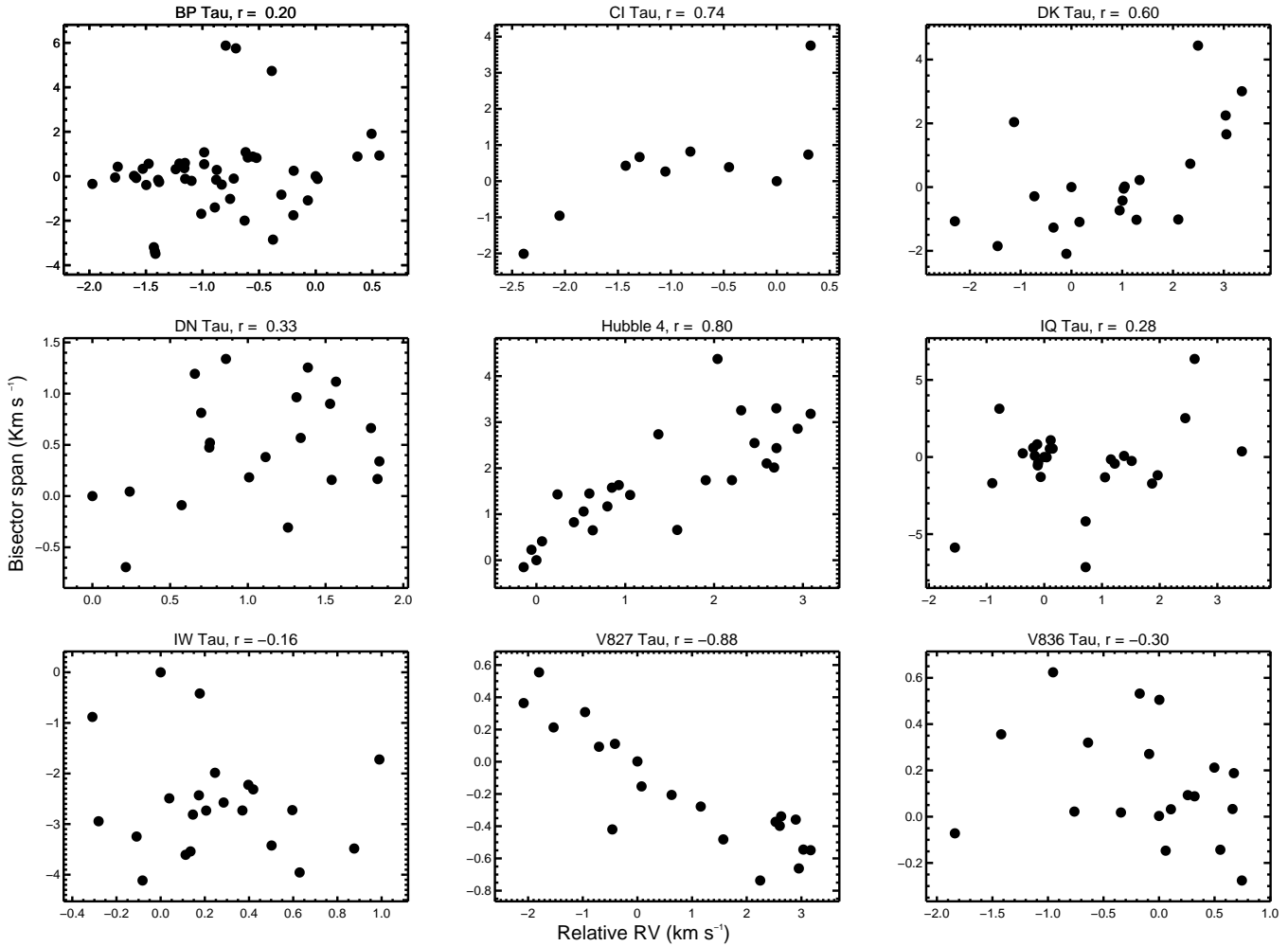}
\caption{Optical RV versus bisector span for Taurus targets with correlation coefficients.  All data were acquired on the McDonald 2.7-m telescope. Strong correlations suggest that the optical RV modulation is spot-driven.}
\label{fig:rvbis}
\end{figure*}

We analyzed the visible light bisectors to determine the origin of each target's RV variability \citep{Huerta:08:472, Mahmud:11:123}. We used the cross-correlation function from each order (\S\ref{sec:VLRVs}) to determine a bisector span (the inverse of the mean slope of the bisector); the mean across all orders provided a mean bisector span for each observation. We then computed the Pearson linear correlation coefficent, $r$, between the bisector spans and RVs for each target (Figure \ref{fig:rvbis} and Table \ref{tbl:rvbis}).  Those targets with no correlation, except a few ``spotted standard stars'' (\S\ref{sec:taurus_disc}), were selected for NIR follow up observations on CSHELL.

\begin{deluxetable}{lcc}
\tablecaption{RV standards}
\tablewidth{0pt}
\tablehead{ 
Name & $N_{obs}$ & $\sigma_v$ (\mps) } 
\startdata
\cutinhead{CSHELL}
GJ 281    & 48 & 66 \\
HD 65277  &  9 & 36 \\
HD 219538 & 13 & 55 \\
HD 225261 & 14 & 61 \\
\cutinhead{NIRSPEC}
GJ 281    &  9 & 30 \\
HD 10476  &  5 & 129 \\
\enddata
\label{tbl:rv_standards}
\end{deluxetable}

\subsection{NIR radial velocities}
\label{sec:NIRRVs}
All CSHELL and NIRSPEC observations were processed using the same pipeline \crop.  The data were dark-subtracted, flat-fielded, and optimally extracted using an IDL implementation of \citet{Horne:86:609}.  We calculated the RVs using a spectral modeling technique.  This technique uses two high-resolution template spectra to model the stellar spectrum and the telluric features.  For the stellar spectrum, we generated NextGen stellar atmosphere models \citep{Hauschildt:99:377} tailored to the \teff, $\log g$, and metallicity of our targets.  We then used SYNTHMAG \citep{Piskunov:99:515} to create template spectra from the NextGen models along with input from atomic \citep{Kupka:00:590} and CO \citep{Goorvitch:94:535} line lists.  

The telluric absorption features in the K-band provide an absolute wavelength and instrumental profile reference, similar in concept to the iodine gas cell technique used in high-precision optical RV exoplanet surveys \citep{Butler:96:500}.  Using the atmosphere as a ``gas cell" lets us superimpose a relatively stable wavelength reference, which follows the same optical path as the light from the science target, onto our spectra.  This helps alleviate uncertainties introduced by variable slit illumination, changing optical path lengths, etc. We modeled the telluric features using the NOAO telluric absorption spectrum \citep{Livingston:91}.  

In order to match the model to our observations, we applied a number of transformations to our templates: a velocity shift and a power law scaling factor, rotational and instrumental broadening, a second order continuum normalization, and a second order wavelength dispersion.  The composite model was then binned down to the resolution of our CSHELL (or NIRSPEC) data.  We used an IDL implementation of the Levenberg-Marquardt non-linear least squares fitting algorithm\footnote{\texttt{MPFIT}, written by Craig Markwardt} to determine the values of the model parameters which best reproduced our observations.  

We used Monte Carlo techniques to estimate the errors in the model parameters.  For each observation, we generated 100 simulated observations based on the measured noise.  We fit a model to each of the simulated observations which provided us with 100 sets of parameters.  We then calculated the mean and standard deviation of the 100 results for each parameter which we took to be the final best-fit value and uncertainty, respectively.

\begin{figure}
\includegraphics[scale=0.7]{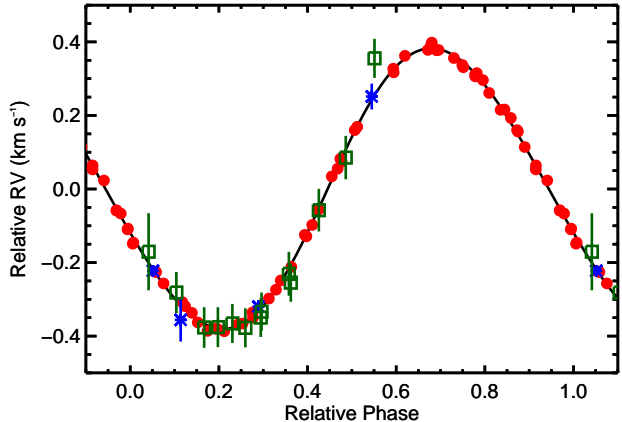}
\caption[NIR RVs for Gl 86 (CSHELL)]{RVs for known exoplanet Gl 86 b.  Filled red circles are CORALIE RVs from \citet{Queloz:00:99}, blue stars are CRIRES RVs from \citep{Figueira:10:55}, and open green squares are our K band CSHELL RVs from Nov 2008 to Nov 2011.  The CSHELL RVs have been corrected for a long term linear drift (see Fig \ref{fig:gl86drift}).  The black line is an orbit model fit to the combined datasets.  The standard deviation of our residuals is 36 \mps.}
\label{fig:gl86}
\end{figure}

\begin{figure}
\includegraphics[scale=0.7]{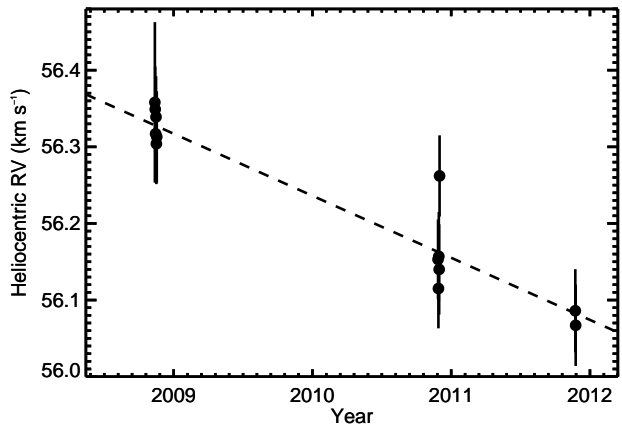}
\caption{Gl 86 CSHELL RVs after subtracting the planet-induced modulation.  The dashed line is a linear fit to the residuals of -0.22 \msd. A similar long term trend was presented by \citet{Queloz:00:99} and indicates the presence of a second, more distant, companion.}
\label{fig:gl86drift}
\end{figure}

The RV for each nod position is the velocity shift of the stellar template relative to the telluric-based wavelength solution.  A final, nightly RV was determined by calculating the average of the individual nod RVs, weighted by their uncertainties.  The final uncertainity in the nightly RV was computed by taking the weighted standard deviation of the nod RVs and dividing by the square root of the number of nods.

To assess the long-term uncertainties in our observations, we routinely observed stars known to have stable RVs (i.e. $\sigma_v <$ 50 \mps).   We observed one RV standard, GJ 281, on every available night to ensure we had a consistent test spanning the entire project. When time allowed, we observed additional RV standards (HD 65277, HD 219538, HD 225261).  Table \ref{tbl:rv_standards} summarizes the RMS RVs for each standard.  We use the RMS of GJ 281, 66 \mps,  as the canonical value of our long-term CSHELL stability since it is the only target observed consistently on all runs.

GJ 281 was also observed on most nights with NIRSPEC, in addition to HD 10476 (Table \ref{tbl:rv_standards}).  We use the RMS of GJ 281, 30 \mps, to quantify our night-to-night uncertainties because, as with CSHELL, it is the most frequently observed standard with the longest time baseline.  Using GJ 281 to assess our NIRSPEC uncertainties also provides a common benchmark for comparisons to our CSHELL observations.  

\section{RV Analysis}
\label{sec:rvanalysis}
\subsection{Gl 86}
\label{sec:gl86}
In \crot, we demonstrated the ability of CSHELL to detect giant planets by confirming the exoplanet companion around Gl 86 \citep{Queloz:00:99, Figueira:10:55}. Figure \ref{fig:gl86} plots new CSHELL RVs with those from \citeauthor{Queloz:00:99} and \citeauthor{Figueira:10:55} along with an orbital fit to all datasets.  Our error bars are determined by adding the uncertainties from our RV algorithm in quadrature with the 66 \mps~systematic uncertainty (\S\ref{sec:NIRRVs}). Our best fit orbit (Table \ref{tbl:gl86}) is very close to the \citeauthor{Queloz:00:99} solution. Our data show good agreement with previously published values exhibiting a $\sim$36 \mps\ standard deviation in the $O-C$ residuals.

\begin{deluxetable}{ccc}
\tablecaption{Gl 86 b parameters}
\tablewidth{0pt}
\tablehead{
Parameter & Value & Units }
\startdata
$K$ & 381.4 & \mps \\
$P$ & 15.764 & days \\
$e$ & 0.05 & \nodata \\
$T_0$ & 2451146.836 & JD \\
$\omega$ & 273.4 & deg \\
$\frac{dv}{dt}$ & -0.22 & m s$^{-1}$ d$^{-1}$ \\
$O-C_{\textrm{\scriptsize{Queloz}}}$ & 7.9 & \mps \\
$O-C_{\textrm{\scriptsize{Figueria}}}$ & 24.2 & \mps \\
$O-C_{\textrm{\scriptsize{Crockett}}}$ & 36.1 & \mps \\
\enddata
\label{tbl:gl86}
\end{deluxetable}

\begin{deluxetable*}{lcccccc}
\tablecaption{Optical and NIR RV amplitudes, standard deviation, and ratios between wavelengths}
\tablewidth{400pt}
\tablehead{
\multicolumn{1}{c}{} & \multicolumn{3}{c}{Amplitudes} & \multicolumn{3}{c}{Standard Deviations} \\
\colhead{Target} & \colhead{Optical} & \colhead{NIR} & \colhead{Ratio} & \colhead{Optical} & \colhead{NIR} & \colhead{Ratio} }
\startdata
BP Tau   &  2.68 $\pm$ 0.26 & 1.70 $\pm$ 0.26 & 1.45 $\pm$ 0.28 & 0.64 $\pm$ 0.03 & 0.43 $\pm$ 0.04 & 1.43 $\pm$ 0.15 \\
CI Tau   &  2.80 $\pm$ 0.30 & 1.90 $\pm$ 0.17 & 1.41 $\pm$ 0.20 & 0.95 $\pm$ 0.08 & 0.63 $\pm$ 0.05 & 1.50 $\pm$ 0.17 \\
DK Tau   &  5.64 $\pm$ 0.31 & 3.46 $\pm$ 0.24 & 1.50 $\pm$ 0.16 & 1.58 $\pm$ 0.07 & 0.85 $\pm$ 0.05 & 1.80 $\pm$ 0.13 \\
DN Tau   &  2.12 $\pm$ 0.26 & 0.87 $\pm$ 0.13 & 2.27 $\pm$ 0.42 & 0.59 $\pm$ 0.05 & 0.27 $\pm$ 0.03 & 2.13 $\pm$ 0.30 \\
Hubble I 4 &  3.34 $\pm$ 0.22 & 1.51 $\pm$ 0.14 & 2.11 $\pm$ 0.23 & 1.07 $\pm$ 0.04 & 0.41 $\pm$ 0.03 & 2.54 $\pm$ 0.19 \\
IQ Tau   & 4.97 $\pm$ 0.35 & 2.59 $\pm$ 0.54 & 1.75 $\pm$ 0.41 & 1.19 $\pm$ 0.07 & 0.77 $\pm$ 0.11 & 1.48 $\pm$ 0.23 \\
IW Tau   &  1.40 $\pm$ 0.18 & 0.87 $\pm$ 0.20 & 1.44 $\pm$ 0.37 & 0.36 $\pm$ 0.03 & 0.25 $\pm$ 0.04 & 1.36 $\pm$ 0.23 \\
V827 Tau &  5.40 $\pm$ 0.86 & 1.16 $\pm$ 0.31 & 4.24 $\pm$ 1.13 & 1.82 $\pm$ 0.06 & 0.37 $\pm$ 0.09 & 4.63 $\pm$ 1.02 \\
V836 Tau &  2.90 $\pm$ 0.42 & 1.06 $\pm$ 0.30 & 2.45 $\pm$ 0.74 & 0.80 $\pm$ 0.08 & 0.35 $\pm$ 0.09 & 2.11 $\pm$ 0.54 \\
\enddata
\label{tbl:allratios}
\end{deluxetable*}

\begin{figure}[hb!]
\includegraphics[scale=1.0]{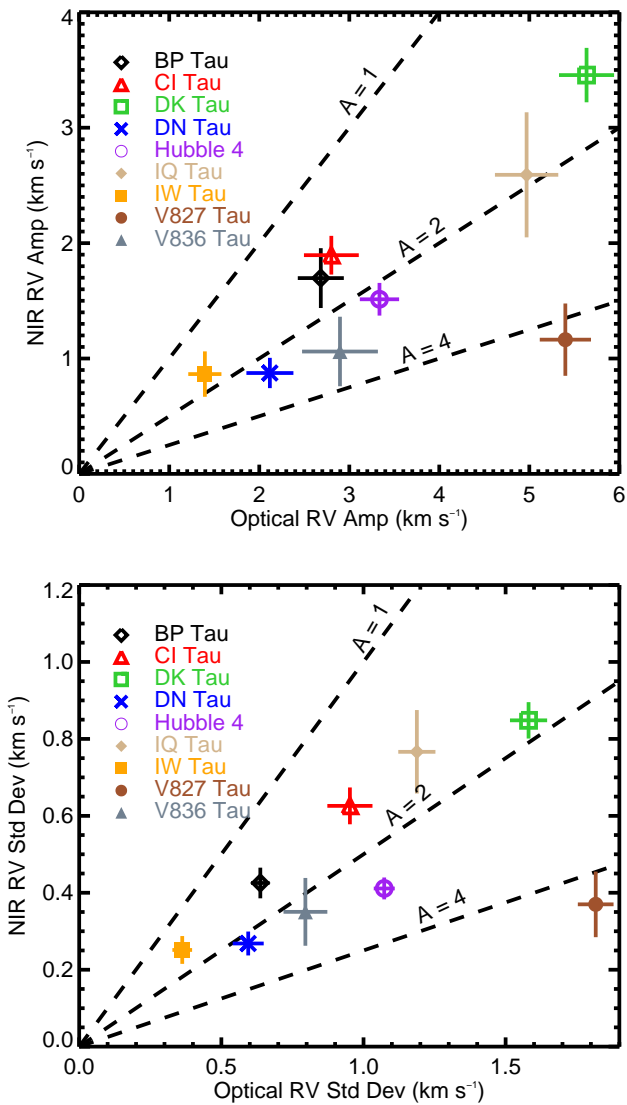}
\caption{Optical and NIR RV amplitudes (top) and standard deviations (bottom).  The dashed lines in each plot show optical--NIR ratios of one, two, and four.  Targets with primarily planet-induced modulation should fall along the line $A=1$ whereas those with larger $A$ values have stronger spot signatures.}
\label{fig:allratios}
\end{figure}

The CSHELL RVs in Figure \ref{fig:gl86} have been corrected for a long term linear drift in this object, first reported in \citet{Queloz:00:99}.  Figure \ref{fig:gl86drift} plots our RVs with the planet modulation subtracted and a linear fit to the residuals.  A Monte Carlo analysis shows that our data are consistent with a 0.22 $\pm$ 0.04 \msd\ drift.  This is shallower than the 0.36 \msd\ drift reported by \citeauthor{Queloz:00:99}  Both datasets suggest another companion with an orbital period on the order of hundreds of years ($a > 20$ AU).  Our results demonstrate the ability of CSHELL to detect not only short-period Jupiter-mass planets but also secular trends from more distant companions.

\subsection{Discussion of individual Taurus targets}
\label{sec:taurus_disc}

\subsubsection{Overview}
\label{sec:taurus_overview}

From the 56 stars on our target list with a complete set of twelve visible light observations, only thirteen have met our criteria for NIR follow-up observations.  Eight of those currently have enough high S/N observations to make meaningful comparisons between the two wavelength bands (the remaining five will continue to be monitored in upcoming observing runs).  In addition to these ``planet host candidates", we observed a known spotted star, V827 Tau, to help us understand how spots will affect the long wavelength RV behavior. These targets, along with relevant stellar parameters, are listed in Table \ref{tbl:stellarparameters}.  The remainder of the sample will be discussed in another paper.

For each of our targets, we calculated the peak-to-peak amplitude of the RV variation in both optical and NIR wavelengths, looked for periodicity at each wavelength, and fit Keplerian models to signficant periods in both bands.  To estimate the amplitudes and their associated uncertainties, we generated $10^5$ datasets for every target, replacing each RV with a Gaussian random number; the mean and standard deviation were set to the observed RV and its associated error bar, respectively.  For each simulated dataset, we recorded the peak-to-peak amplitudes for the optical and NIR RVs ($A_O$ and $A_I$, respectively) and the amplitude ratio between the two wavelengths ($R_A$); we did the same for the RV standard deviations in the two spectral bands.  We then fit Gaussian functions to histograms of all measured values.  The mean of each Gaussian was taken to be the canonical value of that parameter with error bars equal to the standard deviation (Table \ref{tbl:allratios}, Figure \ref{fig:allratios}).

To look for periodicity in our RV measurements, we used an IDL implementation of the CLEAN algorithm \citep{Roberts:87:968}.  CLEAN attempts to remove the couplings between physical periods and their aliases by deconvolving the spectral window function from the discrete Fourier transform.  We calculated the power spectrum for each target using the RVs from the a) optical observations only, b) NIR observations only, and c) combined optical and NIR datasets. 

\begin{figure*}
\includegraphics{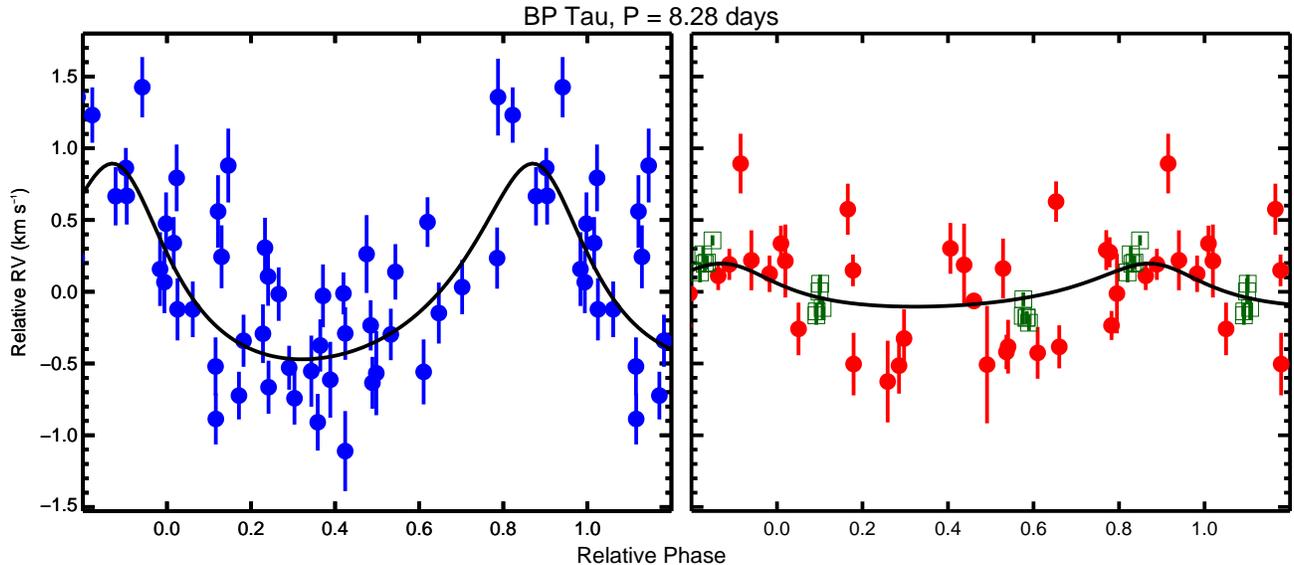}
\caption{Optical RVs (left) and NIR RVs (right) of BP Tau phased to 8.28 days.  On the right, red circles are CSHELL data, green squares are NIRSPEC. The black line is a best-fit Kepler model.  The large change in amplitude between the optical and NIR Kepler models indicates that the RV modulation is most likely spot-induced.}
\label{fig:keplerrv_bptau_8.28}
\end{figure*}

\begin{figure*}
\includegraphics{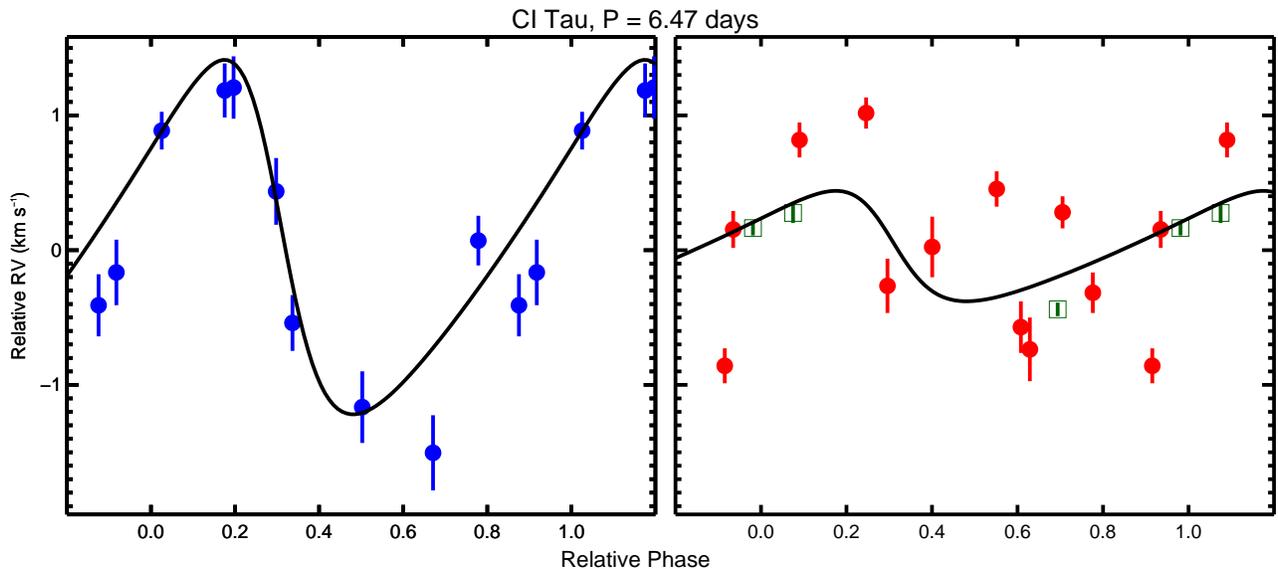}
\caption{Optical RVs (left) and NIR RVs (right) of CI Tau phased to 6.47 days. On the right, red circles are CSHELL data, green squares are NIRSPEC. The black line is a best-fit Kepler model. The extremely poor fit to the NIR data indicates that the periodicity seen in the optical RVs is not evident in the NIR.  The NIR and optical RVs are most likely caused by different physical mechanisms (see Fig.~\ref{fig:keplerrv_citau_10.87}).}
\label{fig:keplerrv_citau_6.47}
\end{figure*}

We also fit a Keplerian model to both the optical and NIR data for each target phased to significant periods found in the periodicity analysis.  To help us quantify differences in the optical and NIR RVs, we started by fitting five free parameters ($K$, $e$, $T_0$, $\omega$, and $V_0$) to the optical data.  The period was fixed to the strongest peak in the optical power spectrum.  Since we were primarily interested in how the RV amplitude changes at the longer wavelengths, we repeated the fit to the NIR data while allowing only the amplitude to vary.  All other parameters were fixed at the best fit values for the optical RVs.  For targets which showed strong periodicity in the NIR, we performed the additional step of reversing this procedure: we fit a model to the NIR RVs with the period fixed to the peak of the NIR power spectrum.  We then fit that model to the optical data, once again allowing only the amplitude to vary.  For clarity, we refer to these four models as follows: optical Kepler fit to the optical data (OKOD), optical Kepler fit to the infrared data (OKID), infrared Kepler fit to the infrared data (IKID), and infrared Kepler fit to the optical data (IKOD).

\begin{figure*}
\includegraphics{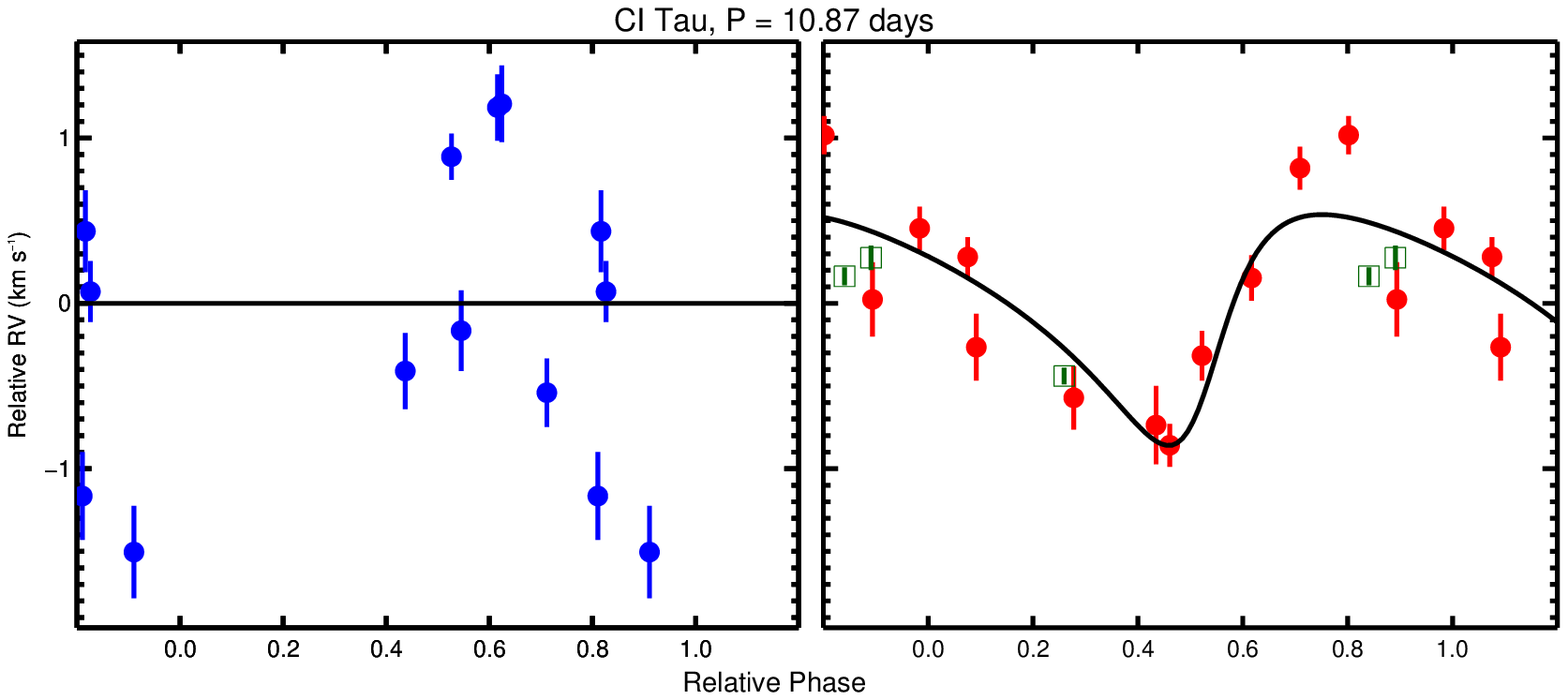}
\caption{Same as Fig.~\ref{fig:keplerrv_citau_6.47} but phased to the strongest NIR period at 10.87 days. A Kepler model is unable to fit the optical data at this period (black line, left) but does provide a good fit to the NIR RVs.  This, coupled with the 6.47 day fit and optical bisector analysis, suggest that the optical RVs are predominately spot-driven whereas the NIR RVs arise from a different mechanism, possibly a companion.}
\label{fig:keplerrv_citau_10.87}
\end{figure*}

\begin{figure*}
\includegraphics{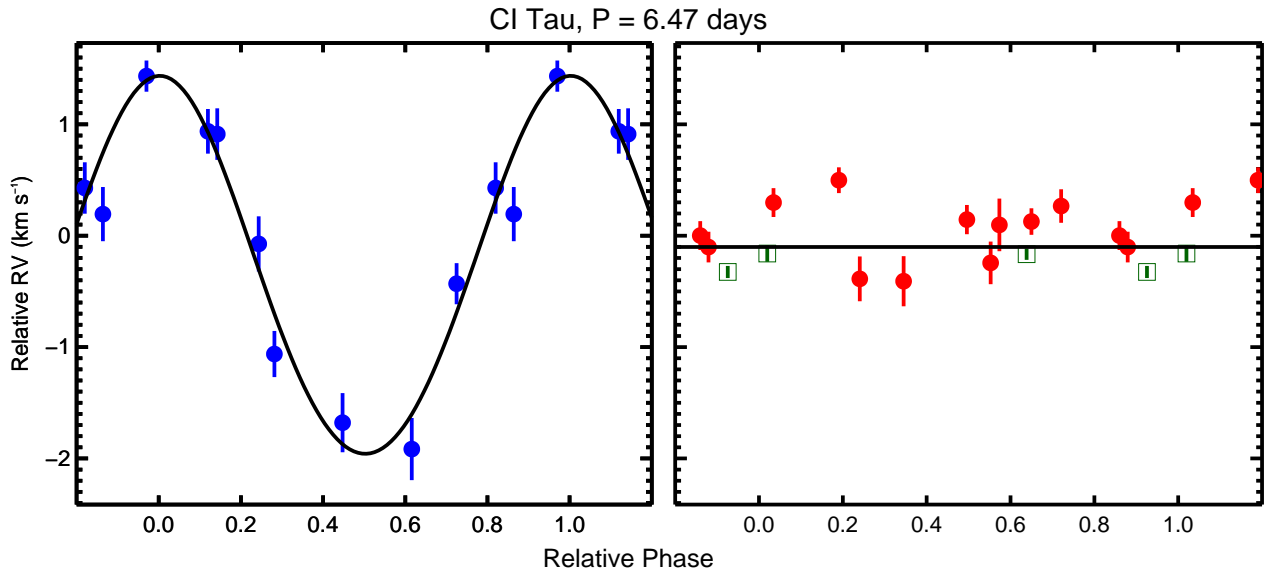}
\caption{Optical RVs (left) and NIR RVs (right) of CI Tau after the best fit NIR Kepler model (Figure \ref{fig:keplerrv_citau_10.87}) has been subtracted from both data sets.  On the right, red circles are CSHELL data, green squares are NIRSPEC. The black line is a new Kepler model fit to the modified data.  The optical RV modulation is much cleaner than Fig.~\ref{fig:keplerrv_citau_6.47} and the NIR RV residuals, as expected, show little variation.  This supports our suggestion that the optical RV modulation is the superposition of both spot and companion induced variability.}
\label{fig:keplerrv_citau_noplanet}
\end{figure*}

\begin{figure*}
\includegraphics{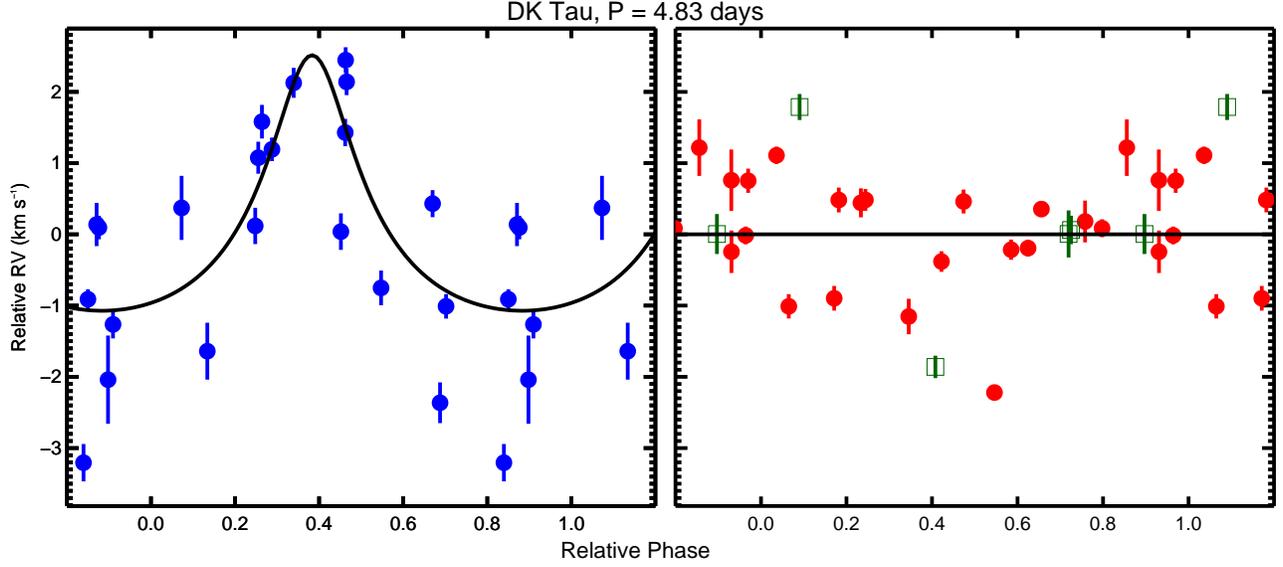}
\caption{Optical RVs (left) and NIR RVs (right) of DK Tau phased to 4.83 days. On the right, red circles are CSHELL data, green squares are NIRSPEC. The black line is a best-fit Kepler model.  Despite being the strongest period in the optical periodogram, the fit is poor at both wavelengths.}
\label{fig:keplerrv_dktau_4.83}
\end{figure*}

\begin{figure*}
\includegraphics{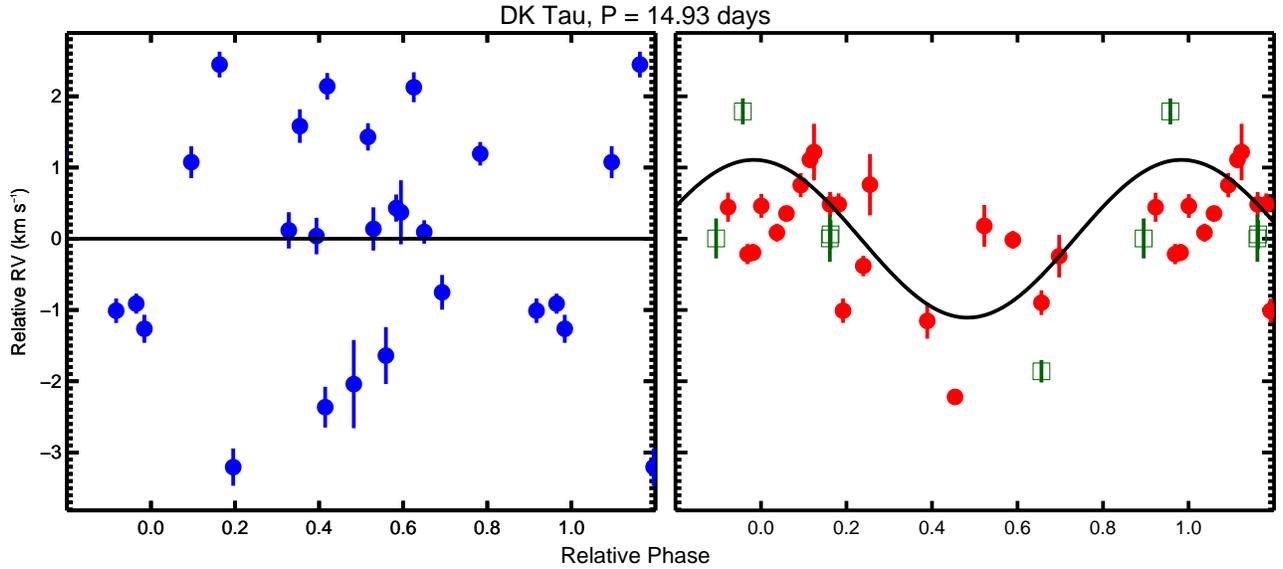}
\caption{Optical RVs (left) and NIR RVs (right) of DK Tau phased to 14.93 days. On the right, red circles are CSHELL data, green squares are NIRSPEC. The black line is a best-fit Kepler model.  The optical RVs show no periodicity with this phasing (as evidenced by the attempt at Kepler fit).  The NIR RV fit is much better than the one seen at 4.83 days, though with considerable scatter.  This may be indicative, like CI Tau, of different physical mechanisms driving the optical and NIR RV modulation.}
\label{fig:keplerrv_dktau_14.93}
\end{figure*}

\begin{figure*}
\includegraphics{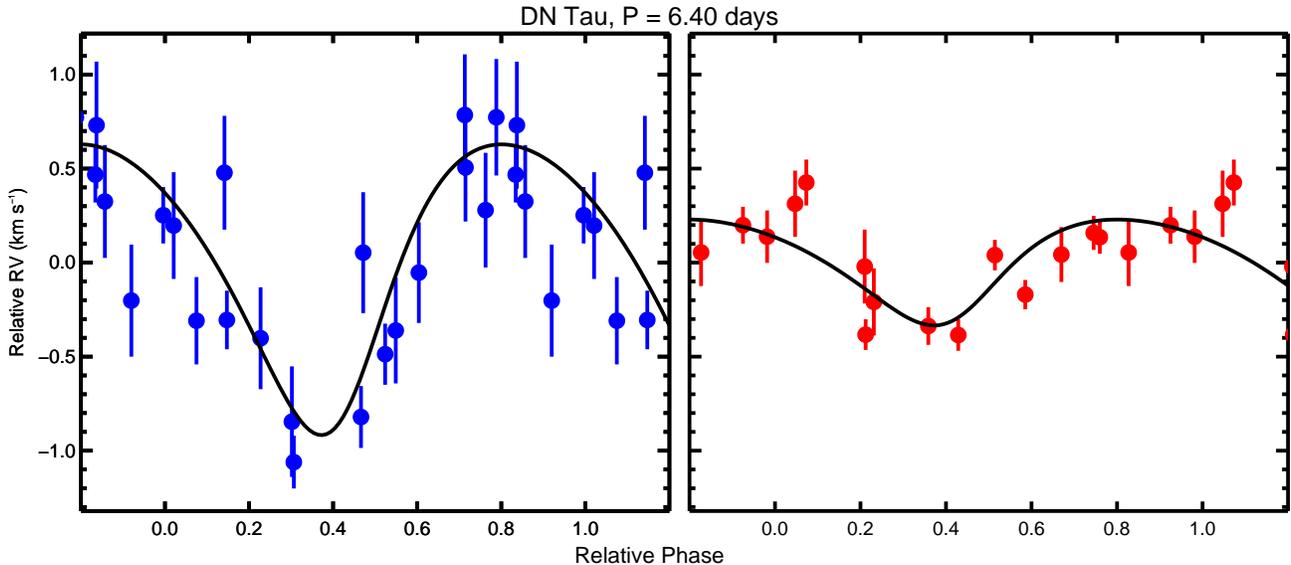}
\caption{Optical RVs (left) and CSHELL RVs (right) of DN Tau phased to 6.40 days. The black line is a best-fit Kepler model.  The reduced amplitude of the NIR fit indicates that this RV variability is primarily spot-induced.}
\label{fig:keplerrv_dntau_6.40}
\end{figure*}

\begin{figure*}
\includegraphics{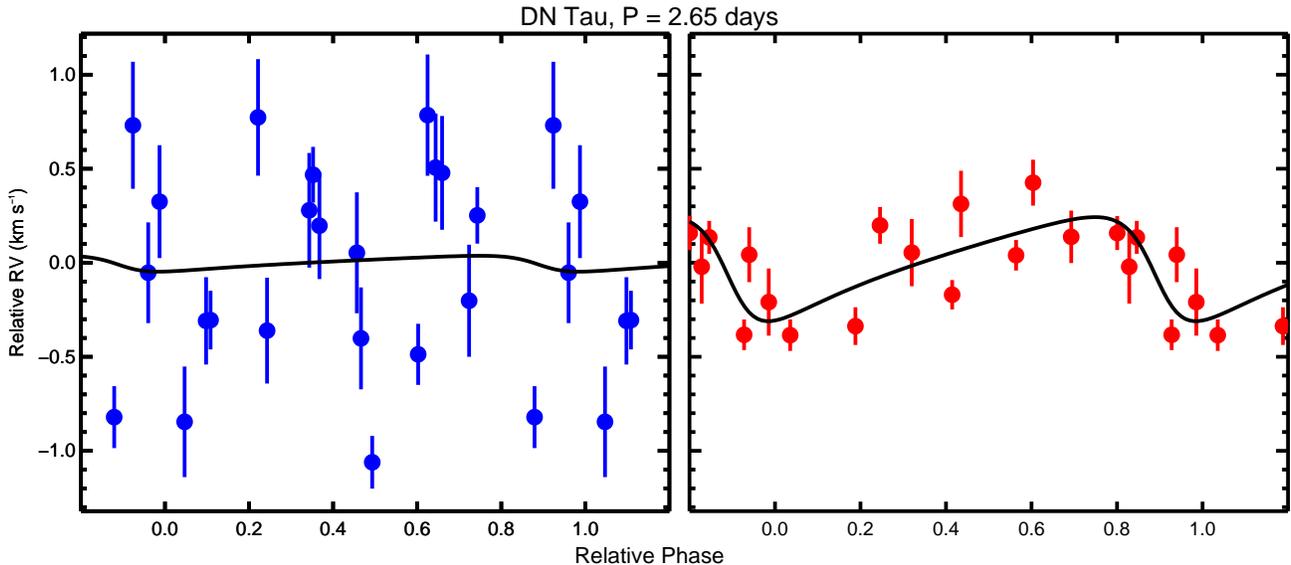}
\caption{Optical RVs (left) and CSHELL RVs (right) of DN Tau phased to 2.65 days. The black line is a best-fit Kepler model.  This period is the strongest seen in the NIR periodogram.  The optical data do not phase to this period.  The NIR fit hints at a different mechanism driving the NIR variability but further investigation is needed.}
\label{fig:keplerrv_dntau_2.65}
\end{figure*}

\begin{figure*}
\includegraphics{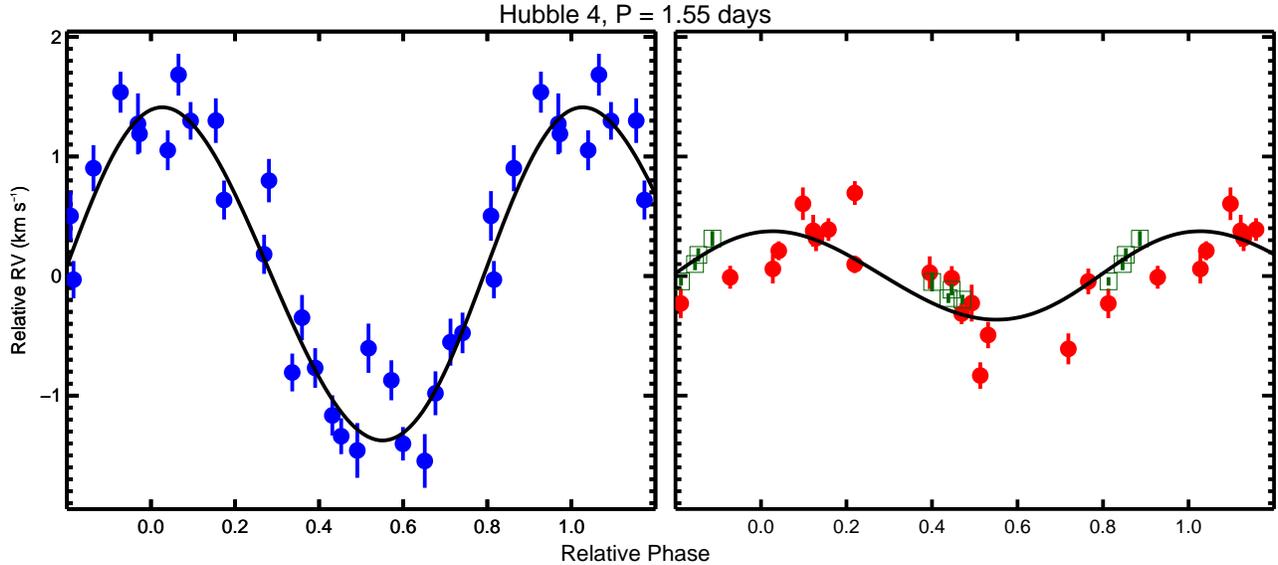}
\caption{Optical RVs (left) and NIR RVs (right) of Hubble I 4 phased to 1.55 days.  On the right, red circles are CSHELL data, green squares are NIRSPEC. The black line is a best-fit Kepler model.  The sharply reduced amplitude in the NIR and consistent phasing between the two wavelengths is a strong argument for spot-driven RV modulation.}
\label{fig:keplerrv_hubble4}
\end{figure*}

\begin{figure*}
\includegraphics{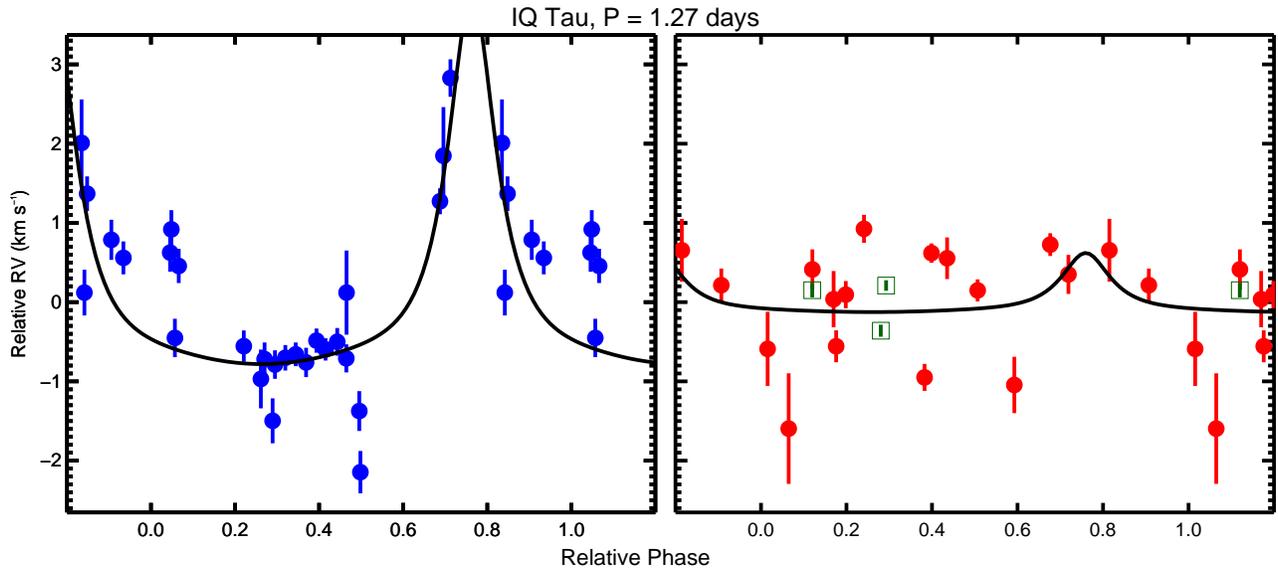}
\caption{Optical RVs (left) and NIR RVs (right) of IQ Tau phased to 1.27 days.  On the right, red circles are CSHELL data, green squares are NIRSPEC. The black line is a best-fit Kepler model.  The NIR RVs show little variability outside their error bars, indicating that the optical RV variability is most likely spot-induced.}
\label{fig:keplerrv_iqtau_1.27}
\end{figure*}

\begin{figure*}
\includegraphics{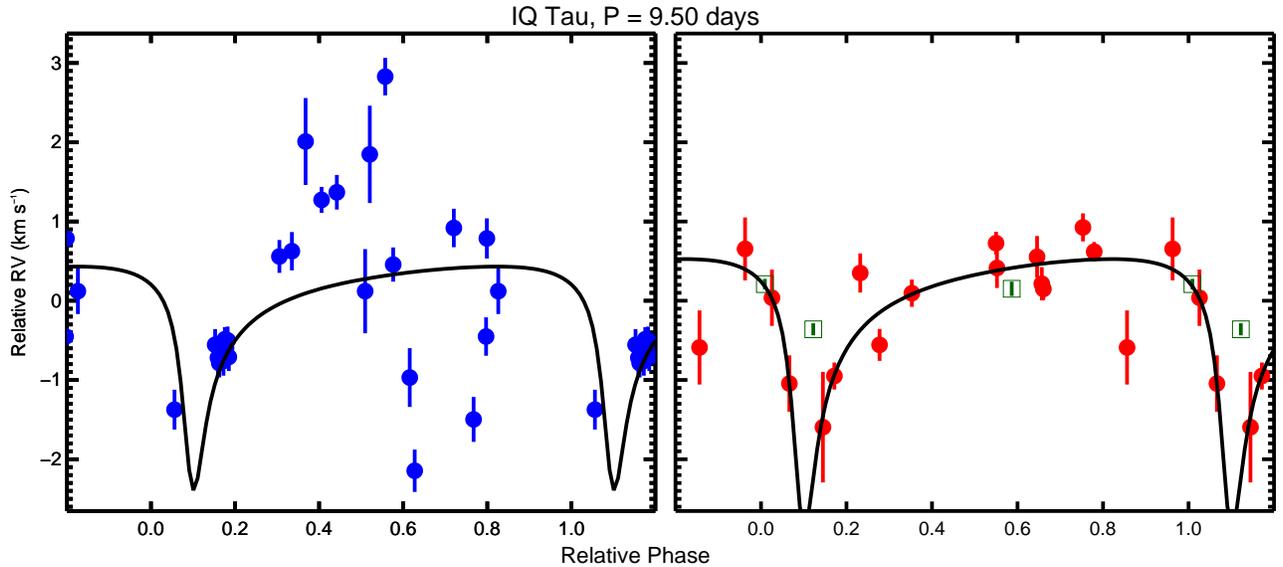}
\caption{Optical RVs (left) and NIR RVs (right) of IQ Tau phased to 9.50 days. On the right, red circles are CSHELL data, green squares are NIRSPEC. The black line is a best-fit Kepler model. The 9.5 day period is the strongest seen in the NIR periodogram.  The optical RVs show considerable scatter around the model, most likely caused by star spots.  The NIR fit is driven mostly by the CSHELL RVs; the NIRSPEC RVs do not show the same degree of variability.}
\label{fig:keplerrv_iqtau_9.50}
\end{figure*}

\begin{figure*}
\includegraphics{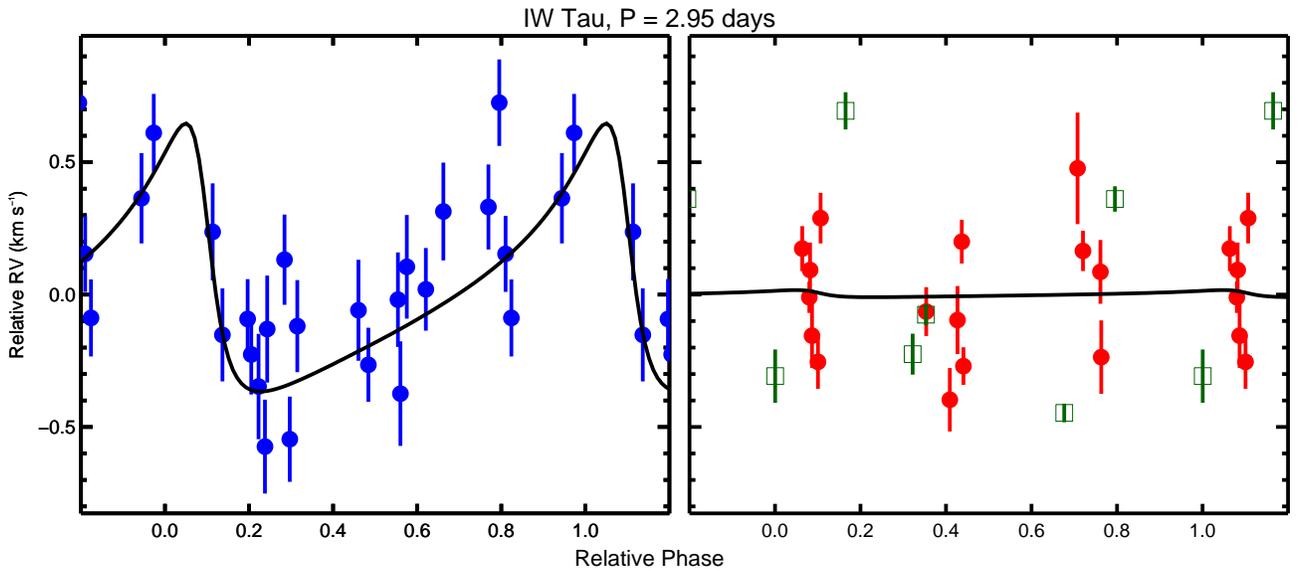}
\caption{Optical RVs (left) and NIR RVs (right) of IW Tau phased to 2.95 days. On the right, red circles are CSHELL data, green squares are NIRSPEC. The black line is a best-fit Kepler model.  The extremely poor fit on the right shows that the strongest optical period does not show itself in the NIR data.  There may be a different physical mechanism behind the large variability seen in the two wavelengths.}
\label{fig:keplerrv_iwtau_2.95}
\end{figure*}

\begin{figure*}
\includegraphics{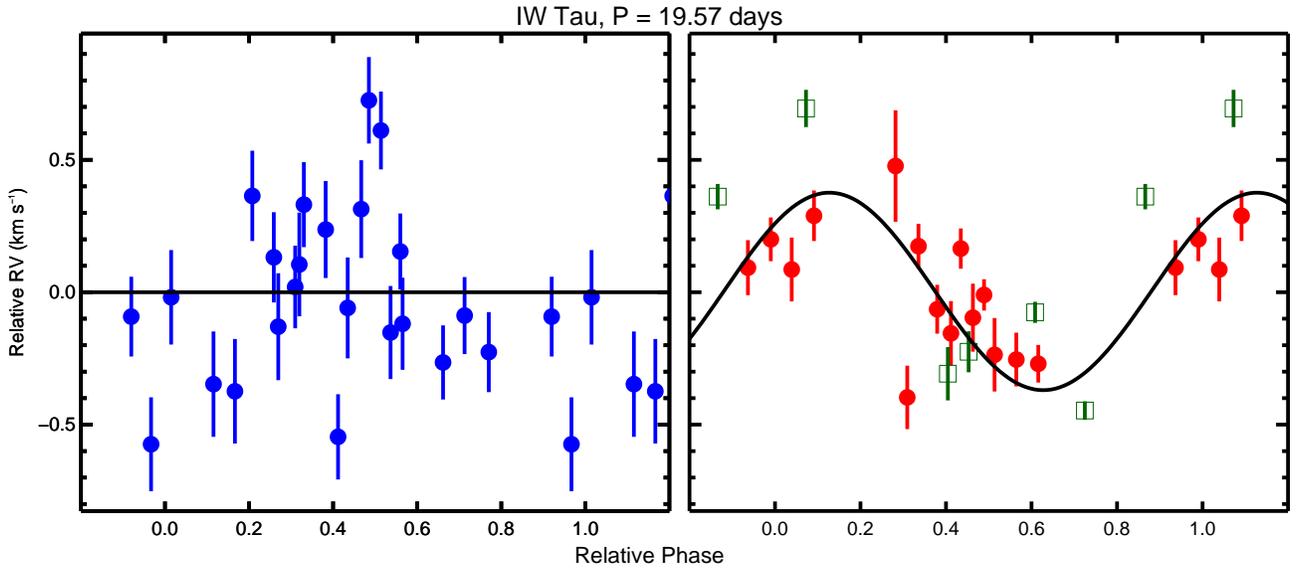}
\caption{Optical RVs (left) and NIR RVs (right) of IW Tau phased to 19.57 days. On the right, red circles are CSHELL data, green squares are NIRSPEC. The black line is a best-fit Kepler model.  At this period, the optical RVs do not phase nor do they fit a Keplerian model.  The NIR RVs exhibit much better phase coherence than that seen at 2.95 days.  There is, however, large scatter seen in the NIR residuals.}
\label{fig:keplerrv_iwtau_19.57}
\end{figure*}

\begin{figure*}
\includegraphics{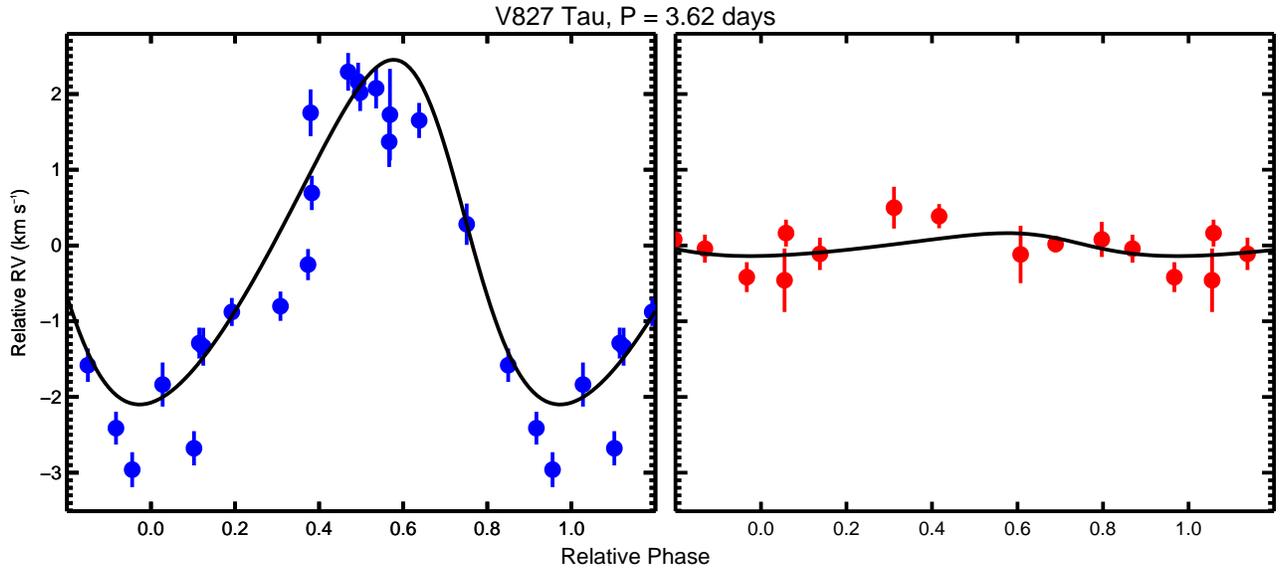}
\caption{Optical RVs (left) and CSHELL RVs (right) of V827 Tau - our ``spotted standard'' - phased to 3.62 days.   The black line is a best-fit Kepler model.  The drastic change in amplitude between the two wavelengths is the classic signature of a heavily spotted star.}
\label{fig:keplerrv_v827}
\end{figure*}

\begin{figure*}
\includegraphics{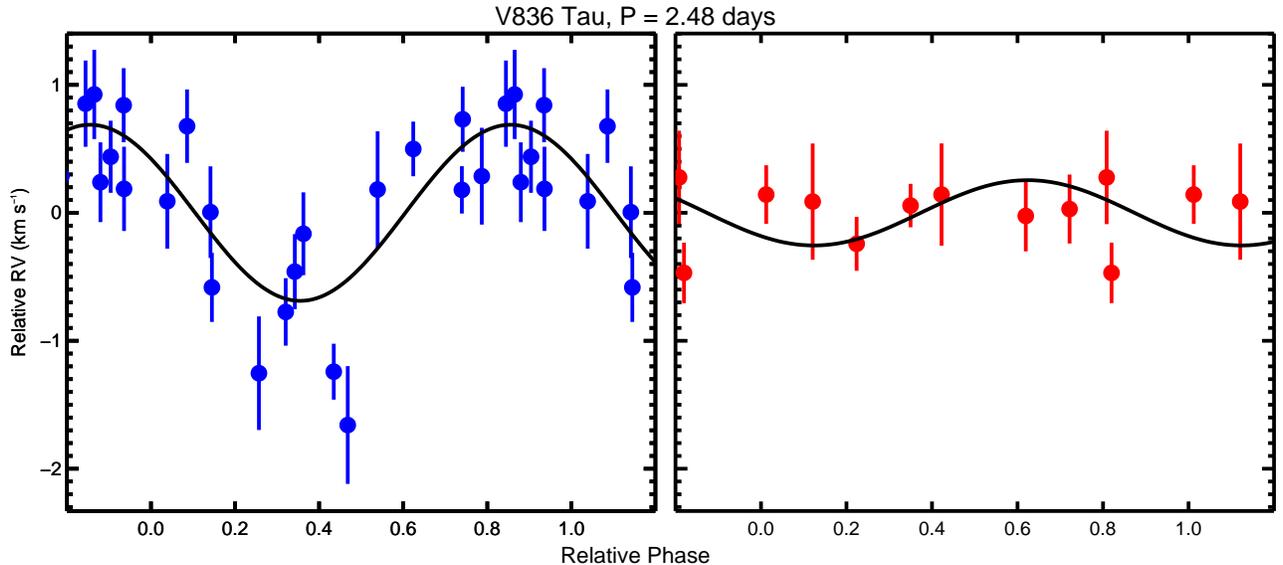}
\caption{Optical RVs (left) and CSHELL RVs (right) of V836 Tau phased to 2.48 days. The black line is a best-fit Kepler model.  There is little variability outside the error bars in the NIR RVs, suggesting that this target's RV modulation is spot-induced.}
\label{fig:keplerrv_v836_2.48}
\end{figure*}

\begin{figure*}
\includegraphics{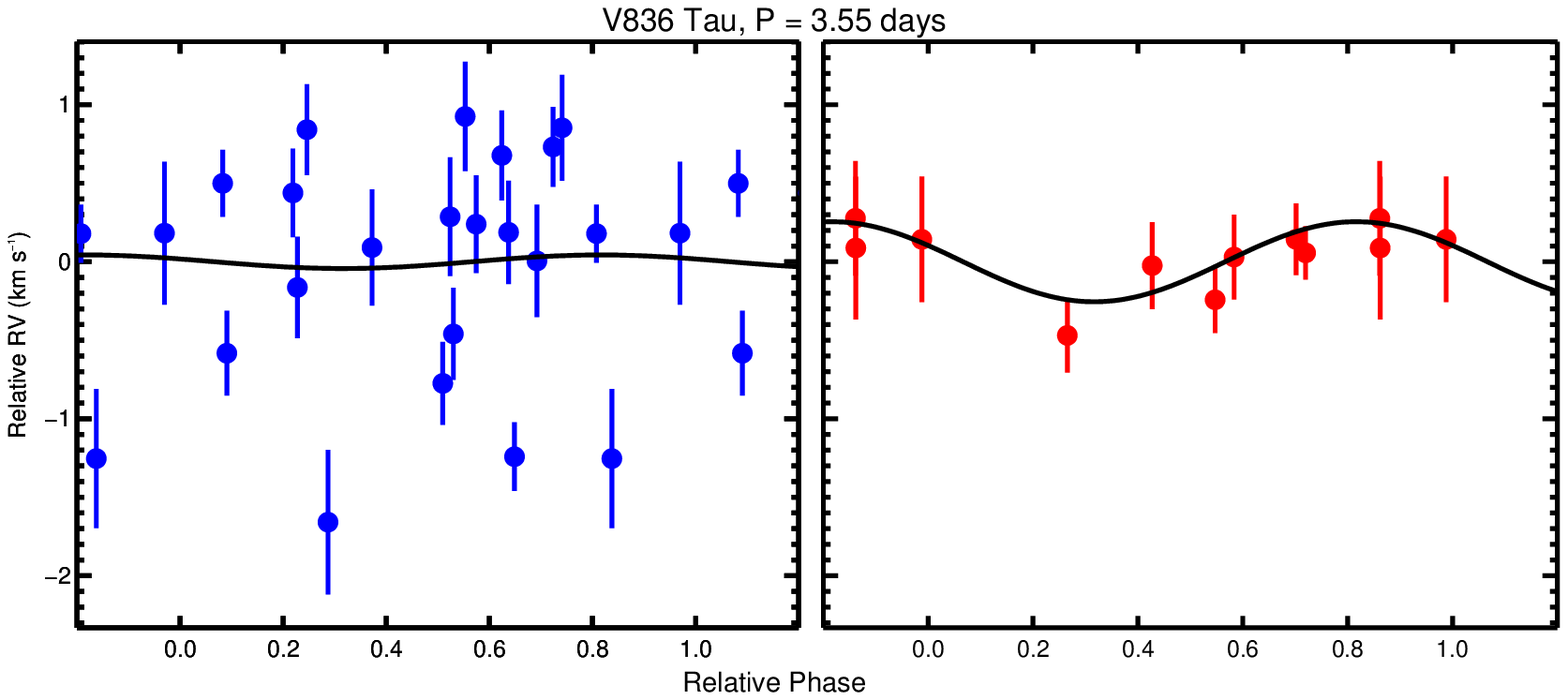}
\caption{Optical RVs (left) and CSHELL RVs (right) of V836 Tau phased to 3.55 days. The black line is a best-fit Kepler model.  Phased to a longer period, the optical RVs lose phase coherence whereas the NIR RVs show a slightly better fit to a Kepler model.  The amplitude is, however, at the same scale as the error bars.  The reasonable NIR fit suggests that higher precision RV follow-up (i.e., with NIRSPEC) is warrented before the companion hypothesis can be ruled out.}
\label{fig:keplerrv_v836_3.55}
\end{figure*}

\subsubsection{BP Tau}
\label{sec:bptau}
BP Tau is our most intensively observed target.  Initial NIR observations strongly supported a giant planet hypothesis; continued observing has failed to uphold that.  BP Tau exhibits significant optical RV variability with $A_O = 2.7$ \kmps\ and a period of 8.28 days; the optical--NIR peak-to-peak RV amplitude ratio ($R_A$) is 1.45.  The Kepler fits to the phase-folded RV curves (OKOD and OKID, Figure \ref{fig:keplerrv_bptau_8.28}) suggest a spot-driven interpretation of the RV modulation.  The ratio of optical-NIR Kepler model amplitudes ($R_K$) is $\sim$4.5.  The NIRSPEC data, in particular, show only slight variation at about the 2$\sigma$ level.   

The optical RV period closely matches photometeric periods reported in the literature of 8.3 and 8.19 days  \citep{Richter:92:1144, Percy:06:1390}.  High time-resolution photometric data from the Trans-Atlantic Exoplanet Survey (TrES) also show a comparable period of 8.309 days \citep{Xiao:12}.  However, periods of 6.1, 7.6, and 7.7 days have also been identified \citep[][respectively]{Simon:90:1957,Vrba:86:199,Osterloh:96:267}.  The correlation with known photometric periods, the ratio of both peak-to-peak and model amplitudes, as well as the lack of any other significant periodicity in the NIR RVs, all point to BP Tau's observed, short-period RV variation being spot-induced.

\subsubsection{CI Tau}
\label{sec:citau}

CI Tau exhibits strong optical RV periodicity at 6.47 days.  The NIR data do not phase meaningfully to this period but do exhibit significant variability ($R_A = 1.41$, Figure \ref{fig:keplerrv_citau_6.47}).  In the combined CSHELL and NIRSPEC data, we find a peak in the periodogram at 10.87 days which is not seen in the optical data. Figure \ref{fig:keplerrv_citau_10.87} presents the same optical and NIR RVs phased to this other period.   In this case, we freely fit all parameters in the NIR Kepler fit and then fit the optical data by fixing all parameters except the amplitude.  The optical fit is clearly nonsensical; there is no periodicity at 10.87 days in the optical RVs.  

Assuming the IKID models a true companion, we subtracted the planet modulation from both the optical and NIR RVs.  Figure \ref{fig:keplerrv_citau_noplanet} illustrates the modified RVs phased to the 6.47 day period.  The optical RVs demonstrate significant improvement.  A Keplerian model to the new optical RVs also shows an excellent fit.  In this scenario, the optical Keplerian parameters are most likely modeling the spot modulation.   The same model fit to the subtracted NIR RVs, allowing only the amplitude to vary, is flat.  

Although for CI Tau we have fewer than twelve optical observations, we see a strong correlation ($r$ = 0.74) between the optical RVs and bisector spans.  Furthermore, we observed significant variability in the K band with a different periodicity than the optical RVs.  These data suggest that a heavily spotted star with a planet may exhibit optical variability that is entirely consistent with a spot-only model for the RV modulation.  \emph{Large spots may mask planet-induced Doppler shifts.}  Modeling is needed to verify this possibilty (Crockett et al.\ in prep).

\subsubsection{DK Tau}
\label{sec:dktau}
DK Tau exhibits large RV amplitudes in both wavelengths: 5.6 \kmps\ in the optical and 3.5 \kmps\ in the NIR. An amplitude ratio of $R_A = 1.5$ and a linear correlation coefficient of $r = 0.6$ are consistent with spot-induced RV variability.  It is not clear, however, that a simple spot model is sufficient to explain the observed behavior.  Figure \ref{fig:keplerrv_dktau_4.83} shows the optical and NIR RVs phase-folded to the strongest period in the optical power spectrum, $P$ = 4.83 days.  There is significant scatter in the optical RVs. After subtracting the OKOD, the standard deviation in the residuals is $\sim$1 \kmps.  Furthermore, the NIR RVs do not phase to this period at all; the amplitude of the OKID is flat with $\sim$1 \kmps\ residuals.

The strongest period in the NIR RVs is, as for CI Tau (\S\ref{sec:citau}), significantly different than that seen in the optical: 14.93 days. Figure \ref{fig:keplerrv_dktau_14.93} shows RV data from both wavelengths phased to this period.  The IKID is a better fit ($\sigma_{O-C}$ $\sim$ 760 \mps) than the OKID, but the NIR RVs still show considerable scatter.  The optical data do not phase to this period.

\subsubsection{DN Tau}
\label{sec:dntau}
DN Tau was interpreted as a spotted star in \prat\ on the basis of eight NIR RVs obtained in February 2008.  Revisiting these data with our improved NIR reduction pipeline hinted at a possibly more complex system so DN Tau was added back to the CSHELL target list for more intensive follow up.  DN Tau shows little RV-bisector correlation ($r$ = 0.33) and considerable scatter in the NIR ($A_I \approx 870$ \mps).  The amplitude ratio ($R_A\approx2.3$), however, seems to again place this target firmly in the spot-induced category.  Figure \ref{fig:keplerrv_dntau_6.40} shows the optical and NIR data both phased to 6.4 days (the strongest peak in the optical periodogram) along with the OKOD and OKID.  The ratio of Keplerian model amplitudes ($R_K \approx 2.7$) further supports the spot hypothesis.

However, interpretation of DN Tau is, similar to CI Tau (\S\ref{sec:citau}), muddied by a second signficant period in the NIR data at 2.65 days (Figure \ref{fig:keplerrv_dntau_2.65}).  To fit the NIR data, we followed the same approach taken for CI Tau.  The IKOD confirms the optical power spectrum analysis: there is no periodicity at 2.65 days in the optical RVs.  This is another target that will benefit from increased NIR scrutiny.

\subsubsection{Hubble I 4}
\label{sec:hubble4}

\citet{Mahmud:11:123} reported that the RV modulation of Hubble I 4 is consistent with a large polar spot on a nearly pole-on star.  Here we refine that analysis with additional NIRSPEC RVs and improvements to the NIR spectrum modeling (Figure \ref{fig:keplerrv_hubble4}).

We previously published a rotation period of 1.5459 days arguing that the slightly stronger 2.81 day period---seen in a Scargle periodogram analysis---was most likely an alias of the true period.  We based this argument on an observed reduction in the 2.81 day power in the CLEAN periodogram and on a published SuperWASP photometric period of 1.5483 days \citep{Norton:07:785}.  
\citet{Xiao:12} also report a photometeric period of 1.547 days seen in TrES data (private communication). The addition of eight NIRSPEC observations have strengthened the case for a 1.5459 day period.  Revisiting the periodogram analysis of \maht\ with this new data, the 2.81 day period no longer appears in the power spectrum.

\subsubsection{IW Tau}
\label{sec:iwtau}
IW Tau is another target that has comparable RV amplitudes in optical and NIR wavelengths ($R_A \approx 1.4$) and shows evidence of different periods in the two bands.  At the strongest optical period (2.95 days), the optical RVs fit reasonably well to an eccentric Kepler model whereas the NIR RVs do not phase to this period at all (Figure \ref{fig:keplerrv_iwtau_2.95}).  The strongest NIR period is 19.57 days (Figure \ref{fig:keplerrv_iwtau_19.57}).  Phased to this period, the optical RVs show little periodicity while the NIR RVs, though plagued by significant scatter, show an adequate fit to a Keplerian model.

Not only does Hubble I 4 support our argument that optical observations are not sufficient when looking for planets around young, active stars, it argues that relying solely on NIR observations is also not conclusive.  Hubble I 4 was the first target in our survey to exhibit resolved RV modulation in the K band.  These data alone are consistent with the presence of a giant planet.  However, comparison to the much larger variability in the optical data, with the same periodicity and phase, demonstrate that the RV modulation of Hubble I 4 is activity-induced.
 
\subsubsection{IQ Tau}
\label{sec:iqtau}
At a typical V magnitude of 14.5, IQ Tau is one of our faintest McDonald targets.  The optical RVs show periodicity at 1.27 days with $A_O \approx 5$ \kmps.  The NIR RVs do not phase to this period and have a lower peak-to-peak amplitude of 2.6 \kmps (Figure \ref{fig:keplerrv_iqtau_1.27}).  The reduction in RV variability at longer wavelengths is strong evidence that the optical RV behavior is spot-induced. The NIR power spectrum does have a peak at 9.5 days, shown in Figure \ref{fig:keplerrv_iqtau_9.50}.  However, the IKID seems to be biased towards a slight dip from one data point at phase 0.1.  The NIRSPEC RVs are not consistent with this drop and, in fact, show very little variation overall.  

During our observing runs at McDonald in early 2012, imaging at the 0.8-meter revealed dramatic variations in magnitude.  Four months of intensive optical photometery from McDonald Observatory, Lowell Observatory, and the U.S. Naval Observatory Flagstaff Station (to be be published in a later paper) show extreme periodic ($\sim$6 days) BVRI magnitude variations up to 2.5 mags.  We have not yet determined any reasonable explanation for the photometeric variability of this system.

\subsubsection{V827 Tau}
\label{sec:v827tau}
We added V827 Tau to our target list as a ``spotted standard star" \prap.  We have used it as a benchmark for how a known spotted star should behave at both wavelengths.   We see strong periodicity in the optical RVs at 3.62 days.  Optical RVs and bisector span show a strong correlation (Figure \ref{fig:rvbis}) with a linear correlation coefficient of $r = -0.88$, consistent with spot-driven RV modulation. Figures \ref{fig:allratios} and \ref{fig:keplerrv_v827} show a clean separation between the optical and NIR ranges with $R_A \approx 4.2$.  The Keplerian fits in Figure \ref{fig:keplerrv_v827} also show a stark contrast.  The OKOD has an amplitude of 2.28 \kmps, whereas the amplitude for the OKID is only 0.15 \kmps---a ratio of $\sim$15!  

V827 Tau is another excellent demonstration of the need for follow-up of optical observations at longer wavelengths: the optical data alone fit reasonably well to a Keplerian model whereas the NIR data show little variation outside of the uncertainties. Without the NIR data (and bisector analysis), it would be easy to conclude that the RV variability was planet-induced.

\subsubsection{V836 Tau}
\label{sec:v836tau}
V836 Tau was also originally presented in \citet{Prato:08:L103} with the conclusion that, despite a lack of RV-bisector correlation ($r$ = -0.3), the RV modulation was spot-induced and not the result of an unseen companion.  The numbers presented here are updated based on additional NIR observations and improvements to the CSHELL data reduction pipeline. Despite significant optical RV variability ($A_O$ = 2.9 \kmps), the NIR scatter is almost entirely within the CSHELL error bars.  With $R_A \approx 2.45$ and $R_K \approx 2.7$ ($P$ = 2.48 days, Figure \ref{fig:keplerrv_v836_2.48}), the data support the spot hypothesis.

We do see some periodicity in the NIR data at 3.55 days (Figure \ref{fig:keplerrv_v836_3.55}).  We repeated the approach of CI Tau and DN Tau by fixing the optical model parameters to those determined by the NIR fit (\S\ref{sec:dntau}).  The IKID fits the NIR data better than the OKID; the infrared model does not fit the optical RVs at all.  However, the low level of NIR variability prevents us from saying anything definitive.  This is yet another target that can greatly benefit from intensive NIR follow-up, particularly with NIRSPEC.

\section{Discussion}
\label{sec:discussion}
Our initial hypothesis that spot-induced RV modulation should decrease at NIR wavelengths was primarily based on multiwavelength photometric surveys \citep[e.g.,][]{Carpenter:01:3160,Vrba:86:199}.  However, there was a lack of long term NIR spectroscopic monitoring of T Tauri stars on which to rely.  Our own spot simulations \citep[to be published in a later paper, but see also][]{Huerta:08:472, Mahmud:11:123}, in addition to other published models \citep{Desort:07:983, Reiners:10:432, Ma:12:172}, predict that NIR RV monitoring reduces the stellar jitter of young stars by a factor of 2--3.  Data from several young stars---TW Hya \citep{Huelamo:08:L9}, V827 Tau \citep{Prato:08:L103}, and Hubble I 4 \citep{Mahmud:11:123}---confirm this prediction.  Furthermore, observations of Hubble I 4 demonstrate another important point: \emph{NIR observations alone are not sufficient}.  Even in the K band, some young stars exhibit RV amplitudes on the order of hundreds of meters per second with good phase coherence over several years. Only by comparing amplitudes in \emph{at least} two widely separated bands can we hope to distinguish between stellar activity and true companions.

Observations, however, have forced us to consider that a cool spot model, at times, may be too simplistic.  BP Tau, for example, exhibits high levels of NIR RV variability ($\sim$1.7 \kmps\ peak-to-peak) which does not phase to the optical data, nor shows any strong periodicity of its own, and can not be explained solely by the presence of star spots. \citet{Eiroa:02:1038} present results from simultaneous optical and NIR photometry of PMS stars which suggest alternate explanations.  They divide their sample into two groups: (i) stars with similar variability trends in optical and NIR bands and (ii) stars with different trends.  Those stars in (i) can be explained by the presence of star spots or variable obscuration (such as from a dusty disk).  Stars in (ii) are more challenging and not all show the expected decrease in amplitude variation with wavelength.  In some cases the NIR variability is greater than that in the optical; other targets get brighter in one band while getting dimmer in another.  This suggests that for some PMS stars, the optical and NIR variability have different underlying physical mechanisms.  While the optical light is dominated by the photosphere, the NIR flux has some contribution from the circumstellar disk.  They estimate that the disk contributes more than 50\% of the K band flux in their sample and thus conclude that the NIR variability of stars with a NIR excess predominately tracks changes in the disk.  

In the case of BP Tau, \citet{Folha:99:517} calculate a K band veiling of 0.8 $\pm$ 0.3 based on high resolution spectroscopy near the 2.1661 \micron\ Br$\gamma$ line.  This implies that roughly 44\% of the K band light is non-photospheric.  We therefore can not rule out the possibility that our K band observations are impacted by either changes in the disk structure or shocks from the base of magnetospheric accretion columns.  \citet{Muzerolle:03:L149} present models which support this hypothesis.  They show that the NIR excess in a sample of classical T Tauri stars (including BP Tau) is consistent with a single-temperature blackbody with $T \sim 1400$ K.  The amount of excess also roughly correlates with accretion luminosity.  Furthermore, \citet{Najita:07:369} report variability in BP Tau's short wavelength NIR excess.  \citet{Dutrey:03:1003} suggest that BP Tau may represent a class of transition object in the process of clearing out its disk based on anomalously low $^{12}$CO to dust emission ratios.  \citeauthor{Dutrey:03:1003}'s imaging of CO emission raises the possibility that the CO photospheric lines used to measure the stellar RV may be contaminated by variable CO emission caused by asymmetries in a warm, rotating disk.  A disk or accretion component to the spectrum modeling may therefore be necessary to fully disentangle the various possible sources of K band variability.  In future work, we plan to combine extant photometric data from the McDonald Observatory 0.8 meter with our spectra to help clarify the underlying source of RV variability in not only BP Tau but all of our targets.

CI Tau, DN Tau, and possibly V836 Tau also illustrate the importance of multiwavelength observations in the case of heavily spotted stars.  In these cases, the optical spot noise completely masks what may be smaller RV variability from an orbiting planet.  CI Tau is particularly interesting in this regard: the strong correlation between optical RVs and bisector spans ($r$ = 0.7) leads to the conclusion that the optical modulation is entirely activity induced.  Nevertheless, we see strong NIR variability \emph{at a different period than that seen in the optical data}.  Furthermore, subtracting the NIR variability from the optical RVs reveals a significantly cleaner modulation.  In this case, we have a target where the optical RVs are almost entirely spot-dominated while the NIR RV variations have a different origin.  Here, our approach of filtering targets by their optical activity fails.  A far more preferable approach is to start with NIR RV monitoring.  Unfortunately, we are limited by practicality.  A lack of community-accessible, efficient, high-resolution NIR spectrographs dictates our emphasis on optical RVs for target selection.  

Additionally, we tested the significance of the periods for all targets with Monte Carlo simulations.  In general, we found that the targets with the greatest number of observations (e.g., BP Tau, Hubble I 4) have reasonably low ($\lesssim$ 6\%) false alarm probabilities (FAP).  For most other targets, the low number of observations combined with a high degree of variability make the periods less certain.  To more firmly establish the true periods, and eventually determine orbital parameters for likely companions, many more high-cadence NIR observations are needed.

\section{Conclusions}
\label{sec:conclusions}
Consistent monitoring of RV standards and known exoplanet hosts has demonstrated the long term viability of our techniques at a time when NIR RV observations are coming into their own.  With a stability of 66 \mps\ over four years, over 20\% of the extant exoplanet population is accessible to our observing methodology.  Furthermore, this success has been demonstrated on a 3-meter class telescope using a twenty year old Cassegrain-mounted spectrograph.  While parallel efforts on large telescopes (Keck, VLT) are highly encouraging, success on a small telescope, where there is greater community access, is arguably of even greater value.  The large blocks of consecutive nights required for a project like this are very difficult to acquire on 8--10 m class telescopes.  This work supports the findings of the recent ReSTAR (Renewing Small Telescopes for Astronomical Research) report: small ($<$ 6 meters) telescopes continue to produce innovative science that not only complements efforts on 8--10 m class facilities but also explores niches for which large telescopes are not well-suited.

While we have demonstrated the capabilities of our techniques, we have yet to confirm the presence of any planets in our young star sample. \citet{Mayor:11} report that 65\% of Sun-like stars harbor a planet of any mass with $P < 100$ days.  Of the known RV-detected exoplanets with $P < 100$ days (such that their orbits are well sampled in our survey), 4\%\ have semi-amplitudes greater than the 200 \mps\ needed for a robust 3$\sigma$ detection with CSHELL\footnote{\url{http://exoplanets.org}}.  Since the T Tauri stars of today will evolve to become the solar analogues of tomorrow, we expect 3--4 planet detections out of the entire 143 star sample ($143 \times 0.65 \times 0.04$) \emph{if} the pre--main sequence planet population mirrors the main sequence one.\footnote{These numbers appear sensitive to selection criteria.  Broadening our criteria to planets with any RV data (including RV followup of transit detections) as opposed to limiting the sample to those systems discovered by RV surveys increases the expected number of detections to 8--9 planets.} Currently, only CI Tau is a strong contender; the periodicity seen in the other two candidates, DN Tau and V836 Tau, is so far only suggestive.

The lack of planet detections to date, while likely impacted by small number statistics, may indicate a real difference between the main sequence and T Tauri planet populations. In situ formation of hot Jupiters is unlikely given both the unreasonable requirements for delivery of solids to the inner disk and accretion efficiency.  Therefore, short-period massive planets most likely arrive at their final orbits via migration \citep[e.g.,][]{Kley:12}.  If hot Jupiters are not present around T Tauri stars, this can place a lower limit of 1--3 Myr on the migration timescale. In this scenario, the Jupiters have formed, most likely beyond the snowline, but have not yet migrated into short-period orbits.  If hot Jupiters arrive via a migration mechanism with a timescale greater than several million years, a survey of young star associations over a range of ages would be beneficial.  By identifying the age at which hot Jupiters are detected, such a survey could place observational limits to the migration timescale.

After seven years of observations, we have also yet to discover new short-period stellar or brown dwarf companions.  Similar findings have been uncovered for brown dwarf companions to main sequence stars. While $\sim$16\% of solar analogues have close ($P < 5$ years) companions, less than 1\% of those are brown dwarfs; the remainder are stellar ($\sim$11\%) or giant planet ($\sim$5\%) companions \citep[the ``brown dwarf desert'', e.g.,][]{Grether:06:1051}. Even before completion of our survey, this null result on short-period brown dwarf detections to date suggests that the desert is not evolutionary in nature \citep[e.g.,][]{Armitage:02:L11} but rather intrinsic to the formation of brown dwarfs.  

T Tauri stars reside in complex environments and, as such, require careful observations and analyses to isolate the various origins of their observed variability.  We may find that between star spots, changing effective temperatures, shocks, and disk obscuration, these young stars are too poorly suited to RV analysis.   All other approaches, however, have just as many, if not more, drawbacks.  Direct imaging is the only technique that is beginning to successfully sample the young star population, but it is limited to very massive planets on wide orbits.  Only an RV survey can adequately sample the inner AUs---possibly the terrestrial planet forming region.  The potential reward therefore justifies the intensive effort required.  While the discovery of any one planet is unlikely to discriminate between diverse planet formation models, the confirmation of a planet around any of our targets would be paradigm-shifting.  It would represent the first solid data point of a hot Jupiter from the epoch of active planet formation and provide one of the first observational limits on the formation timescale.

\acknowledgments
The authors thank our anonymous referee for a thorough and enthusiastic review.  We acknowledge the SIM Young Planets Key Project for research support; funding was also provided by NASA Origins Grants 05-SSO05-86 and 07-SSO07-86.  This work made use of the SIMBAD database, the NASA Astrophysics Data System, and the Two Micron All Sky Survey (2MASS), a joint project of the University of Massachusetts and IPAC/Caltech, funded by NASA and the NSF.  Some of the data presented herein were obtained at the W.M. Keck Observatory, which is operated as a scientific partnership among the California Institute of Technology, the University of California and the National Aeronautics and Space Administration. The Observatory was made possible by the generous financial support of the W.M. Keck Foundation. We recognize the significant cultural role that Mauna Kea plays in the indigenous Hawaiian community and are grateful for the opportunity to observe there.

{\it Facilities:} \facility{IRTF (CSHELL)}, \facility{Keck:II (NIRSPEC)}, \facility{Smith (Tull)}

\bibliography{pub,unpub}

\begin{thebibliography}{68}
\expandafter\ifx\csname natexlab\endcsname\relax\def\natexlab#1{#1}\fi

\bibitem[{Anglada-Escude {et~al.}(2012)Anglada-Escude, Plavchan, Mills, Gao,
  Garcia-Berrios, Lewis, Sung, Ciardi, {et~al.}}]{Anglada-Escude:12}
Anglada-Escude, G., {et~al.} 2012, arXiv

\bibitem[{Armitage \& Bonnell(2002)}]{Armitage:02:L11}
Armitage, P., \& Bonnell, I. 2002, MNRAS, 330, L11

\bibitem[{Bailey {et~al.}(2012)Bailey, White, Blake, Charbonneau, Barman,
  Tanner, \& Torres}]{Bailey:12:16}
Bailey, J.~I., White, R.~J., Blake, C.~H., Charbonneau, D., Barman, T.~S.,
  Tanner, A.~M., \& Torres, G. 2012, ApJ, 749, 16

\bibitem[{Bean {et~al.}(2010)Bean, Seifahrt, Hartman, Nilsson, Wiedemann,
  Reiners, Dreizler, \& Henry}]{Bean:10:410}
Bean, J.~L., Seifahrt, A., Hartman, H., Nilsson, H., Wiedemann, G., Reiners,
  A., Dreizler, S., \& Henry, T.~J. 2010, ApJ, 713, 410

\bibitem[{Blake {et~al.}(2010)Blake, Charbonneau, \& White}]{Blake:10:684}
Blake, C.~H., Charbonneau, D., \& White, R.~J. 2010, ApJ, 723, 684

\bibitem[{Blake {et~al.}(2007)Blake, Charbonneau, White, Marley, \&
  Saumon}]{Blake:07:1198}
Blake, C.~H., Charbonneau, D., White, R.~J., Marley, M.~S., \& Saumon, D. 2007,
  ApJ, 666, 1198

\bibitem[{Butler {et~al.}(1996)Butler, Marcy, Williams, McCarthy, Dosanjh, \&
  Vogt}]{Butler:96:500}
Butler, R.~P., Marcy, G.~W., Williams, E., McCarthy, C., Dosanjh, P., \& Vogt,
  S.~S. 1996, PASP, 108, 500

\bibitem[{Carpenter {et~al.}(2001)Carpenter, Hillenbrand, \&
  Skrutskie}]{Carpenter:01:3160}
Carpenter, J.~M., Hillenbrand, L.~A., \& Skrutskie, M.~F. 2001, AJ, 121, 3160

\bibitem[{Crockett {et~al.}(2011)Crockett, Mahmud, Prato, Johns-Krull, Jaffe,
  \& Beichman}]{Crockett:11:78}
Crockett, C.~J., Mahmud, N.~I., Prato, L., Johns-Krull, C.~M., Jaffe, D.~T., \&
  Beichman, C.~A. 2011, ApJ, 735, 78

\bibitem[{Cumming {et~al.}(1999)Cumming, Marcy, \& Butler}]{Cumming:99:890}
Cumming, A., Marcy, G.~W., \& Butler, R.~P. 1999, ApJ, 526, 890

\bibitem[{Desort {et~al.}(2007)Desort, Lagrange, Galland, Udry, \&
  Mayor}]{Desort:07:983}
Desort, M., Lagrange, A.-M., Galland, F., Udry, S., \& Mayor, M. 2007, A\&A,
  473, 983

\bibitem[{Dutrey {et~al.}(2003)Dutrey, Guilloteau, \& Simon}]{Dutrey:03:1003}
Dutrey, A., Guilloteau, S., \& Simon, M. 2003, A\&A, 402, 1003

\bibitem[{Eiroa {et~al.}(2002)Eiroa, Oudmaijer, Davies, de~Winter, Garz\'{o}n,
  Palacios, Alberdi, Ferlet, {et~al.}}]{Eiroa:02:1038}
Eiroa, C., {et~al.} 2002, A\&A, 384, 1038

\bibitem[{Figueira {et~al.}(2010)Figueira, Pepe, Melo, Santos, Lovis, Mayor,
  Queloz, Smette, {et~al.}}]{Figueira:10:55}
Figueira, P., {et~al.} 2010, A\&A, 511, 55

\bibitem[{Folha \& Emerson(1999)}]{Folha:99:517}
Folha, D. F.~M., \& Emerson, J.~P. 1999, A\&A, 352, 517

\bibitem[{Glebocki \& Gnacinski(2005)}]{Glebocki:05}
Glebocki, R., \& Gnacinski, P. 2005, VizieR On-line Data Catalog, 3244, 0

\bibitem[{Goorvitch(1994)}]{Goorvitch:94:535}
Goorvitch, D. 1994, ApJSS, 95, 535

\bibitem[{Greene {et~al.}(1993)Greene, Tokunaga, Toomey, \&
  Carr}]{Greene:93:313}
Greene, T.~P., Tokunaga, A.~T., Toomey, D.~W., \& Carr, J.~B. 1993, Proc. SPIE,
  1946, 313

\bibitem[{Grether \& Lineweaver(2006)}]{Grether:06:1051}
Grether, D., \& Lineweaver, C.~H. 2006, ApJ, 640, 1051

\bibitem[{Hatzes(2002)}]{Hatzes:02:392}
Hatzes, A.~P. 2002, Astronomische Nachrichten, 323, 392

\bibitem[{Hatzes {et~al.}(1997)Hatzes, Cochran, \& Johns-Krull}]{Hatzes:97:374}
Hatzes, A.~P., Cochran, W.~D., \& Johns-Krull, C.~M. 1997, ApJ, 478, 374

\bibitem[{Hauschildt {et~al.}(1999)Hauschildt, Allard, \&
  Baron}]{Hauschildt:99:377}
Hauschildt, P.~H., Allard, F., \& Baron, E. 1999, ApJ, 512, 377

\bibitem[{Hebrard {et~al.}(2011)Hebrard, Ehrenreich, Bouchy, Delfosse, Moutou,
  Arnold, Boisse, Bonfils, {et~al.}}]{Hebrard:11:L11}
Hebrard, G., {et~al.} 2011, A\&A, 527, L11

\bibitem[{Herbig \& Bell(1988)}]{Herbig:88}
Herbig, G.~H., \& Bell, K.~R. 1988, {Third Catalog of Emission-Line Stars of
  the Orion Population} (Lick Observatory Bulletin)

\bibitem[{Hinkle {et~al.}(2000)Hinkle, Wallace, Valenti, \& Harmer}]{Hinkle:00}
Hinkle, K., Wallace, L., Valenti, J., \& Harmer, D. 2000, {Visible and Near
  Infrared Atlas of the Arcturus Spectrum 3727-9300 A} (San Francisco: ASP)

\bibitem[{Horne(1986)}]{Horne:86:609}
Horne, K. 1986, PASP, 98, 609

\bibitem[{Hu\'{e}lamo {et~al.}(2008)Hu\'{e}lamo, Figueira, Bonfils, Santos,
  Pepe, Gillon, Azevedo, Barman, {et~al.}}]{Huelamo:08:L9}
Hu\'{e}lamo, N., {et~al.} 2008, A\&A, 489, L9

\bibitem[{Huerta {et~al.}(2008)Huerta, Johns-Krull, Prato, Hartigan, \&
  Jaffe}]{Huerta:08:472}
Huerta, M., Johns-Krull, C.~M., Prato, L., Hartigan, P., \& Jaffe, D.~T. 2008,
  ApJ, 678, 472

\bibitem[{Johns-Krull(2007)}]{Johns-Krull:07:975}
Johns-Krull, C.~M. 2007, ApJ, 664, 975

\bibitem[{Johnson {et~al.}(2010)Johnson, Aller, Howard, \&
  Crepp}]{Johnson:10:905}
Johnson, J.~A., Aller, K.~M., Howard, A.~W., \& Crepp, J.~R. 2010,
  $\backslash$pasp, 122, 905

\bibitem[{Kenyon {et~al.}(2008)Kenyon, G\'{o}mez, \& Whitney}]{Kenyon:08:405}
Kenyon, S.~J., G\'{o}mez, M., \& Whitney, B.~A. 2008, {Low Mass Star Formation
  in the Taurus-Auriga Clouds}, Vol.~4 (ASP Monograph Publications), 405

\bibitem[{Kley \& Nelson(2012)}]{Kley:12}
Kley, W., \& Nelson, R.~P. 2012, ArXiv e-prints

\bibitem[{Kraus \& Ireland(2012)}]{Kraus:12:5}
Kraus, A.~L., \& Ireland, M.~J. 2012, ApJ, 745, 5

\bibitem[{Kupka {et~al.}(2000)Kupka, Ryabchikova, Piskunov, Stempels, \&
  Weiss}]{Kupka:00:590}
Kupka, F.~G., Ryabchikova, T.~A., Piskunov, N.~E., Stempels, H.~C., \& Weiss,
  W.~W. 2000, Baltic Astronomy, 9, 590

\bibitem[{Lafreni\`{e}re {et~al.}(2010)Lafreni\`{e}re, Jayawardhana, \& van
  Kerkwijk}]{Lafreniere:10:497}
Lafreni\`{e}re, D., Jayawardhana, R., \& van Kerkwijk, M.~H. 2010, ApJ, 719,
  497

\bibitem[{Livingston \& Wallace(1991)}]{Livingston:91}
Livingston, W., \& Wallace, L. 1991, {An atlas of the solar spectrum in the
  infrared from 1850 to 9000 cm-1 (1.1 to 5.4 micrometer)} (National Solar
  Observatory)

\bibitem[{Ma \& Ge(2012)}]{Ma:12:172}
Ma, B., \& Ge, J. 2012, ApJ, 750, 172

\bibitem[{Mahmud {et~al.}(2011)Mahmud, Crockett, Johns-Krull, Prato, Hartigan,
  Jaffe, \& Beichman}]{Mahmud:11:123}
Mahmud, N.~I., Crockett, C.~J., Johns-Krull, C.~M., Prato, L., Hartigan, P.~M.,
  Jaffe, D.~T., \& Beichman, C.~A. 2011, ApJ, 736, 123

\bibitem[{Mart\'{\i}n {et~al.}(2006)Mart\'{\i}n, Guenther, Osorio, Bouy, \&
  Wainscoat}]{Martin:06:L75}
Mart\'{\i}n, E.~L., Guenther, E., Osorio, M. R.~Z., Bouy, H., \& Wainscoat, R.
  2006, ApJ, 644, L75

\bibitem[{Mayor \& Queloz(1995)}]{Mayor:95:355}
Mayor, M., \& Queloz, D. 1995, Nature, 378, 355

\bibitem[{Mayor {et~al.}(2011)Mayor, Marmier, Lovis, Udry, S\'{e}gransan, Pepe,
  Benz, Bertaux, {et~al.}}]{Mayor:11}
Mayor, M., {et~al.} 2011, arXiv

\bibitem[{McLean {et~al.}(2000)McLean, Graham, Becklin, Figer, Larkin,
  Levenson, \& Teplitz}]{McLean:00:1048}
McLean, I.~S., Graham, J.~R., Becklin, E.~E., Figer, D.~F., Larkin, J.~E.,
  Levenson, N.~A., \& Teplitz, H.~I. 2000, in Society of Photo-Optical
  Instrumentation Engineers (SPIE) Conference Series, Vol. 4008, Society of
  Photo-Optical Instrumentation Engineers (SPIE) Conference Series, ed. M.~Iye
  \& A.~F. Moorwood, 1048--1055

\bibitem[{McLean {et~al.}(1998)McLean, Becklin, Bendiksen, Brims, Canfield,
  Figer, Graham, Hare, {et~al.}}]{McLean:98:566}
McLean, I.~S., {et~al.} 1998, Proc. SPIE, 3354, 566

\bibitem[{Muzerolle {et~al.}(2003)Muzerolle, Calvet, Hartmann, \&
  D'Alessio}]{Muzerolle:03:L149}
Muzerolle, J., Calvet, N., Hartmann, L., \& D'Alessio, P. 2003, ApJL, 597, L149

\bibitem[{Najita {et~al.}(2007)Najita, Strom, \& Muzerolle}]{Najita:07:369}
Najita, J.~R., Strom, S.~E., \& Muzerolle, J. 2007, MNRAS, 378, 369

\bibitem[{Nidever {et~al.}(2002)Nidever, Marcy, Butler, Fischer, \&
  Vogt}]{Nidever:02:503}
Nidever, D.~L., Marcy, G.~W., Butler, R.~P., Fischer, D.~A., \& Vogt, S.~S.
  2002, ApJSS, 141, 503

\bibitem[{Norton {et~al.}(2007)Norton, Wheatley, West, Haswell, Street,
  Cameron, Christian, Clarkson, {et~al.}}]{Norton:07:785}
Norton, A.~J., {et~al.} 2007, A\&A, 467, 785

\bibitem[{Osterloh {et~al.}(1996)Osterloh, Thommes, \& Kania}]{Osterloh:96:267}
Osterloh, M., Thommes, E., \& Kania, U. 1996, A\&AS, 120, 267

\bibitem[{Palla \& Stahler(2002)}]{Palla:02:1194}
Palla, F., \& Stahler, S.~W. 2002, ApJ, 581, 1194

\bibitem[{Paulson {et~al.}(2004)Paulson, Cochran, \& Hatzes}]{Paulson:04:3579}
Paulson, D.~B., Cochran, W.~D., \& Hatzes, A.~P. 2004, AJ, 127, 3579

\bibitem[{Paulson \& Yelda(2006)}]{Paulson:06:706}
Paulson, D.~B., \& Yelda, S. 2006, PASP, 118, 706

\bibitem[{Percy {et~al.}(2006)Percy, Gryc, Wong, \& Herbst}]{Percy:06:1390}
Percy, J.~R., Gryc, W.~K., Wong, J. C.-Y., \& Herbst, W. 2006, PASP, 118, 1390

\bibitem[{Piskunov(1999)}]{Piskunov:99:515}
Piskunov, N. 1999, in Astrophysics and Space Science Library, Vol. 243,
  Polarization, ed. {K. N. Nagendra \& J. O. Stenflo}, 515

\bibitem[{Prato {et~al.}(2008)Prato, Huerta, Johns-Krull, Mahmud, Jaffe, \&
  Hartigan}]{Prato:08:L103}
Prato, L., Huerta, M., Johns-Krull, C.~M., Mahmud, N., Jaffe, D.~T., \&
  Hartigan, P. 2008, ApJ, 687, L103

\bibitem[{Queloz {et~al.}(2000)Queloz, Mayor, Weber, Bl\'{e}cha, Burnet,
  Confino, Naef, Pepe, {et~al.}}]{Queloz:00:99}
Queloz, D., {et~al.} 2000, A\&A, 354, 99

\bibitem[{Reiners {et~al.}(2010)Reiners, Bean, Huber, Dreizler, Seifahrt, \&
  Czesla}]{Reiners:10:432}
Reiners, A., Bean, J.~L., Huber, K.~F., Dreizler, S., Seifahrt, A., \& Czesla,
  S. 2010, ApJ, 710, 432

\bibitem[{Richter {et~al.}(1992)Richter, Basri, Perlmutter, \&
  Pennypacker}]{Richter:92:1144}
Richter, M., Basri, G., Perlmutter, S., \& Pennypacker, C. 1992, PASP, 104,
  1144

\bibitem[{Roberts {et~al.}(1987)Roberts, Lehar, \& Dreher}]{Roberts:87:968}
Roberts, D.~H., Lehar, J., \& Dreher, J.~W. 1987, AJ, 93, 968

\bibitem[{Saar \& Donahue(1997)}]{Saar:97:319}
Saar, S.~H., \& Donahue, R.~A. 1997, ApJ, 485, 319

\bibitem[{Setiawan {et~al.}(2008)Setiawan, Henning, Launhardt, M\"{u}ller,
  Weise, \& K\"{u}rster}]{Setiawan:08:38}
Setiawan, J., Henning, T., Launhardt, R., M\"{u}ller, A., Weise, P., \&
  K\"{u}rster, M. 2008, Nature, 451, 38

\bibitem[{Setiawan {et~al.}(2007)Setiawan, Weise, Henning, Launhardt,
  M\"{u}ller, \& Rodmann}]{Setiawan:07:L145}
Setiawan, J., Weise, P., Henning, T., Launhardt, R., M\"{u}ller, A., \&
  Rodmann, J. 2007, ApJ, 660, L145

\bibitem[{Simon {et~al.}(1990)Simon, Vrba, \& Herbst}]{Simon:90:1957}
Simon, T., Vrba, F.~J., \& Herbst, W. 1990, AJ, 100, 1957

\bibitem[{Tokunaga {et~al.}(1990)Tokunaga, Toomey, Carr, Hall, \&
  Epps}]{Tokunaga:90:131}
Tokunaga, A.~T., Toomey, D.~W., Carr, J., Hall, D. N.~B., \& Epps, H.~W. 1990,
  Proc. SPIE, 1235, 131

\bibitem[{Tull {et~al.}(1995)Tull, MacQueen, Sneden, \& Lambert}]{Tull:95:251}
Tull, R.~G., MacQueen, P.~J., Sneden, C., \& Lambert, D.~L. 1995, PASP, 107,
  251

\bibitem[{van Eyken {et~al.}(2012)van Eyken, Ciardi, von Braun, Kane, Plavchan,
  Bender, Brown, Crepp, {et~al.}}]{vanEyken:12}
van Eyken, J.~C., {et~al.} 2012, ArXiv

\bibitem[{Vrba {et~al.}(1986)Vrba, Rydgren, Chugainov, Shakovskaia, \&
  Zak}]{Vrba:86:199}
Vrba, F.~J., Rydgren, A.~E., Chugainov, P.~F., Shakovskaia, N.~I., \& Zak,
  D.~S. 1986, ApJ, 306, 199

\bibitem[{Wolszczan(1994)}]{Wolszczan:94:538}
Wolszczan, A. 1994, Science, 264, 538

\bibitem[{Xiao {et~al.}(2012)Xiao, Covey, Rebull, Charbonneau, Mandushev,
  O'Donovan, Slesnick, \& Lloyd}]{Xiao:12}
Xiao, H.~Y., Covey, K.~R., Rebull, L., Charbonneau, D., Mandushev, G.,
  O'Donovan, F., Slesnick, C., \& Lloyd, J.~P. 2012, in press

\end{thebibliography}

\end{document}